\documentclass[twocolumn,twocolappendix]{aastex631}
\usepackage{graphicx}
\usepackage{tablefootnote}
\graphicspath{{Figures/}}

\begin{document}

\newcommand{\affilCEA}{Université Paris-Saclay, Université Paris Cité, CEA, CNRS, AIM, F-91191 Gif-sur-Yvette, France}
\newcommand{\affilSTSCI}{Space Telescope Science Institute, 3700 San Martin Drive, Baltimore, MD 21218, USA}
\newcommand{\affilUKATC}{UK Astronomy Technology Centre, Royal Observatory, Blackford Hill, Edinburgh, EH9 3HJ, UK}
\newcommand{\affilSOFIA}{Stratospheric Observatory for Infrared Astronomy, NASA Ames Research Center, Mail Stop 204-14, Moffett Field, CA 94035, USA}
\newcommand{\affilJPL}{Jet Propulsion Laboratory, California Institute of Technology, 4800 Oak Grove Dr., Pasadena, CA 91109, USA}

\title{JWST MIRI Imager observations of Supernova SN~1987A}

\correspondingauthor{P.\ Bouchet}
\email{Patrice.Bouchet@cea.fr}

\author[0000-0002-6018-3393]{P.\ Bouchet}
\affiliation{\affilCEA}

\author[0009-0007-5200-1362]{R.\ Gastaud}
\affiliation{Université Paris-Saclay, CEA, DEDIP, 91191, Gif-sur-Yvette, France}

\author[0000-0001-6492-7719]{A.\ Coulais}
\affiliation{LERMA, Observatoire de Paris, Université PSL, Sorbonne Université, CNRS, Paris, France}
\affiliation{\affilCEA}

\author[0000-0002-3875-1171]{M.\ J.\ Barlow}
\affiliation{Department of Physics and Astronomy, University College
London (UCL), Gower Street, London WC1E 6BT, UK}

\author[0000-0001-8532-3594]{C.\ Fransson}
\affiliation{Department of Astronomy, Stockholm University, The Oskar
Klein Centre, AlbaNova, SE-106 91 Stockholm, Sweden}

\author[0000-0001-6872-2358]{P.\ J.\ Kavanagh}
\affiliation{Department of Experimental Physics, Maynooth University, Maynooth, Co.\ Kildare, Ireland}
\affiliation{Dublin Institute for Advanced Studies, School of Cosmic Physics,
Astronomy \& Astrophysics Section, 31 Fitzwilliam Place, Dublin 2, Ireland.}

\author[0000-0003-0065-2933]{J.\ Larsson}
\affiliation{Department of Physics, KTH Royal Institute of Technology,
The Oskar Klein Centre, AlbaNova, SE-106 91 Stockholm, Sweden}

\author[0000-0001-7380-3144]{T.\ Temim}
\affiliation{Department of Astrophysical Sciences, Princeton University, Princeton, NJ 08544, USA}

\author[0000-0003-4870-5547]{O.\ C.\ Jones}
\affiliation{\affilUKATC}

\author[0000-0002-2954-8622]{A.\ S.\ Hirschauer}
\affiliation{\affilSTSCI}

\author[0009-0003-6128-2347]{T.\ Tikkanen}
\affiliation{School of Physics \& Astronomy, Space Research Centre, University of Leicester, Space Park Leicester, 92 Corporation Road, Leicester LE4 5SP, UK}

\author[0000-0002-5797-2439]{J.\ A.\ D.\ L.\  Blommaert}
\affiliation{Astronomy and Astrophysics Research Group, Department of
Physics and Astrophysics, Vrije Universiteit Brussel, Pleinlaan 2,
B-1050 Brussels, Belgium}

\author[0000-0003-2238-1572]{O.\ D.\ Fox}
\affiliation{\affilSTSCI}

\author[0000-0002-2041-2462]{A.\ Glasse}
\affiliation{\affilUKATC}

\author[0000-0002-2667-1676]{N.\ Habel}
\affiliation{\affilSOFIA}

\author[0000-0002-4571-2306]{J.\ Hjorth}
\affiliation{DARK, Niels Bohr Institute, University of Copenhagen, Jagtvej
128, 2200 Copenhagen, Denmark}

\author[0000-0002-0577-1950]{J.\ Jaspers}
\affil{Department of Experimental Physics, Maynooth University, Maynooth, Co.\ Kildare, Ireland}
\affiliation{Dublin Institute for Advanced Studies, School of Cosmic Physics,
Astronomy \& Astrophysics Section, 31 Fitzwilliam Place, Dublin 2, Ireland.}

\author{O.\ Krause}
\affiliation{Max Planck Institute for Astronomy, Heidelberg (MPIA)}

\author[0000-0003-0778-0321]{R.\ M.\ Lau}
\affiliation{NSF's NOIR Lab 950 N.\ Cherry Avenue, Tucson, AZ 85721, USA}

\author[0000-0003-4023-8657]{L.\ Lenki\'{c}}
\affiliation{\affilSOFIA}

\author[0000-0002-0522-3743]{M.\ Meixner}
\affiliation{\affilJPL}

\author[0000-0001-6576-6339]{O.\ Nayak}
\affiliation{\affilSTSCI}

\author[0000-0002-4410-5387]{A.\ Rest}
\affiliation{\affilSTSCI}
\affiliation{Department of Physics and Astronomy, Johns Hopkins University,
3400 North Charles Street, Baltimore, MD 21218, USA}

\author[0000-0001-9855-8261]{B.\ Sargent}
\affiliation{\affilSTSCI}
\affiliation{Center for Astrophysical Sciences, The William H.\ Miller III Department of Physics and Astronomy, Johns Hopkins University, Baltimore, MD 21218, USA}

\author[0000-0002-4000-4394]{R.\ Wesson}
\affiliation{School of Physics and Astronomy, Cardiff University, Queen's Buildings, The Parade, Cardiff, CF24 3AA, UK}

\author[0000-0001-7416-7936]{G.\ S.\ Wright}
\affiliation{\affilUKATC}

\author[0000-0002-9090-4227]{L.\ Colina}
\affiliation{Centro de Astrobiolog{\'i}a (CAB), CSIC-INTA, Ctra.\ de Ajalvir km 4, Torrej{\'o}n de Ardoz, E-28850, Madrid,, Spain}

\author[0000-0001-7591-1907]{E.\ F.\ Van Dishoeck}
\affiliation{Max-Planck Institut für Extraterresstrische Physic (MPE), Giessenbachstr., 1, D-85748, Garching, Germany}
\affiliation{Leiden Observatory, Leiden University, 2300 RA Leiden, The Netherlands}

%\author[0000-0002-2554-1837]{Th.\ Greve}
%\affiliation{Department of Space Research and %Technology, Astrophysics and Atmospheric Physics, Elektrovej 327, 208, 2800 Kgs.\ Lyngby, Denmark}

\author[0000-0001-9818-0588]{M.\ Güdel}
\affiliation{Dept. of Astrophysics, University of Vienna, Türkenschanzstr. 17, A-1180 Vienna, Austria}
\affiliation{Max-Planck-Institut für Astronomie (MPIA), Königstuhl, 17, D-69117 Heidelberg, Germany}
\affiliation{ETH Zürich, Institute for Particle Physics and Astrophysics, Wolfgang-Pauli Str., 27, 8093 Zürich, Switzerland}

\author[0000-0002-1493-300X]{Th.\ Henning}
\affiliation{Max-Planck Institut für Extraterresstrische Physic (MPE), Giessenbachstr., 1, D-85748, Garching, Germany}

\author{P.-O.\ Lagage}
\affiliation{\affilCEA}

\author[0000-0002-3005-1349]{G.\ Östlin}
\affiliation{Department of Astronomy, Stockholm University, The Oskar
Klein Centre, AlbaNova, SE-106 91 Stockholm, Sweden}

\author[0000-0002-2110-1068]{T.\ P.\ Ray}
\affiliation{Dublin Institute for Advanced Studies, School of Cosmic Physics, Astronomy \& Astrophysics Section 31, Fitzwilliam Place, Dublin 2, Ireland}

\author[0000-0002-1368-3109]{B.\ Vandenbussche}
\affiliation{Institute of Astronomy, KU Leuwen, Celestijnenlaan 200D, 3001 Leuwen, Belgium}

\begin{abstract}

There exist very few mid-infrared (IR) observations of supernovae (SNe) in general.
Therefore, SN~1987A, the closest visible SN in 400 years, gives us the opportunity to explore the mid-IR properties of SNe, the dust in their ejecta and surrounding medium, and to witness the birth of a SN remnant (SNR).
The James Webb Space Telescope (JWST), with its high spatial resolution and extreme sensitivity, gives a new view on these issues.
We report on the first imaging observations obtained with the Mid-InfraRed Instrument (MIRI).
We build temperature maps and discuss the morphology of the nascent SNR.
Our results show that the temperatures in the equatorial ring (ER) are quite non-uniform.
This could be due to dust destruction in some parts of the ring, as had been assumed in some previous works.
We show that the IR emission extends beyond the ER, illustrating the fact that the shock wave has now passed through this ring to affect the circumstellar medium on a larger scale.
Finally, while sub-mm Atacama Large Millimeter Array (ALMA) observations have hinted at the location of the compact remnant of SN~1987A, we note that our MIRI data have found no such evidence.

\end{abstract}


\section{Introduction} % Section 1
\label{sec:intro}

Our interpretation of the Universe is being transformed by the advent of the James Webb Space Telescope \citep[JWST,][]{Gardner2023}.

Its unparalleled infrared (IR) sensitivity provides a revolutionary view into the properties and characteristics of a wide range of astrophysical phenomena.
During the first year of JWST operations, the iconic supernova SN~1987A, the closest optical SN in 400 years (see
\citealp{McCray1993,McCray2016} for reviews) was among the first targets selected for observation.
Astronomers have followed its full evolution across the entire electromagnetic spectrum, owing to its close proximity within the nearby Large Magellanic Cloud (LMC), as it completes its transformation into a SN remnant (SNR).

The first possible spatially-resolved detection of mid-IR emission for any SN were reported by \citet{Bouchet2004}, using the Gemini South 8-m telescope at the position of SN~1987A on day 6067 since the explosion.
Following that, Spitzer observations of SN~1987A spanned more than a decade.
Since day $\sim$4,000 after outburst, its mid-IR emission has been dominated by dust, most probably from the equatorial ring (ER), rather than from the ejecta.
Indeed, the ground-based images of \cite{Bouchet2004} showed this directly, and was confirmed by the lower resolution Spitzer images that required deconvolution (or modelling).
Decomposition of the marginally-resolved emission also confirms mid-IR domination by dust, and shows that the west side has been brightening relative to the other portions of the ER \citep{Bouchet2014}.

From $\sim$6,000 to $\sim$8,000 days after the explosion, Spitzer observations included broadband photometry at 3.6--24 \micron\, and low- and moderate-resolution spectroscopy at 5--35 \micron\ \citep{Arendt2016,Arendt2020}.
After Spitzer's helium cryogen was exhausted in 2009, however, only the IRAC 3.6 and 4.5 \micron\ bands remained operational at the warmer spacecraft temperatures, restricting further IR observations to these imaging wavelengths alone.
While the exact nature of the emission at these wavelengths has not been certain, regular observations of SN~1987A continued in order to monitor the evolution of the interaction of the blast wave with the ER and to develop a clear picture of the evolving relationship between the IR emission at these wavelengths and emission at optical and X-ray wavelengths \citep{Dwek2008,Dwek2010,Arendt2016}.
These data show that the 3.6 and 4.5 \micron\ brightness had clearly begun to fade after day $\sim$8500, and no longer tracked the X-ray emission as well as it had at earlier epochs.
This can be explained by the destruction of the dust in the ER on timescales shorter than the cooling time for the shocked gas.
It was found also that the evolution of the late time IR emission was similar to the fading optical emission at that epoch (see for instance, \citealp{Dwek2010}).

Using the Gemini South 8-m telescope, high-resolution 11.7 and 18.3 \micron\ mid-IR images of SN~1987A were also obtained at day 6,526 \citep{Bouchet2006}.
It was shown that most of the emission arising from the ER was thermal in origin from silicate dust, presumed to have condensed out in the red supergiant wind of the progenitor star.
They estimated the dust temperature to be $\sim$166$\pm$15~K.
%, and the emitting dust mass was $\sim$2.6$\times$10$^{-6}$~M$_{\odot}$.

Comparison of the Gemini 11.7~\micron\ image with Chandra X-ray images, Hubble Space Telescope (HST) UV-optical images, and Australia Telescope Compact Array (ATCA) radio synchrotron images shows generally good correlation across all wavelengths.
If the dust resides in the diffuse X-ray emitting gas then it is collisionally heated.
The IR emission can then be used to derive the plasma temperature and density, which were found to be in good agreement with those inferred from the X-rays \citep{Dwek2010}.
Alternatively, the dust could reside in the dense UV-optical knots and be heated by the radiative shocks that are propagating through the knots. 

Overall, we are now witnessing the interaction of the SN blast wave with its surrounding medium, creating an environment that is rapidly evolving at all wavelengths.
Since its explosion, SN~1987A has evolved from an SN dominated by the emission from the radioactive decay of $^{56}$Co, $^{57}$Co, and $^{44}$Ti in the ejecta, to an SNR whose emission is dominated by the interaction of the SN blast wave with its surrounding medium \citep{Larsson2011}.
The latter consists of an ER flanked by two outer rings \citep{Burrows1995}, possibly part of an hourglass structure \citep{Chevalier1995,Sugerman2005}.

The collision between the ejecta of SN~1987A and the ER predicted to occur sometime in the interval 1995 -- 2007 \citep{Gaensler1997,Borkowski1997} is still underway.
At UV optical (UVO) wavelengths, ``hot spots" have appeared inside the ER \citep{Pun1997}, and their brightness varies on timescales of a few months \citep{Lawrence2000}.
New hot spots have continued to appear as the entire inner rim of the ER has become lit up by the interaction with the blast wave.
HST images with equivalence to $R$-band (WFPC2/F675W, ACS/F625W, and WFC3/F675W; see \citealp{Larsson2021}) obtained between 1994 and 2009 revealed a necklace of such hot spots, nearly filling a lighted ring.
%\citep{Larsson2011}.
Monitoring at X-ray wavelengths with the XMM-Newton, Chandra, and at radio frequencies, showed that while soft X-rays followed the optical and IR evolution, reaching maxima at $8,000 - 10,000$ days, hard X-rays and radio have shown a steady increase in the flux \citep[e.g., Fig.\ 4 in][]{Alp2021}.

Since $\sim$5,000 days post-explosion, bright multi-wavelength emission has been produced by the shock interaction between the ER and ejecta \citep{McCray2016}.
This is observed to currently be fading in IR, optical, and soft X-ray emission \citep{Fransson2015,Larsson2019b,Arendt2020,Alp2021,Maitra2022}, suggesting that shocks are disrupting the dense ER and that the blast wave has passed through it \citep{Fransson2015}.
The free expansion of the dense inner ejecta has simultaneously continued within the ER, revealing a highly-asymmetric distribution in progressively greater detail.

The recent JWST data has provided us with unprecedented information on the mid-IR properties of this system.
The Mid-InfraRed Instrument (MIRI) onboard JWST provides imaging and spectroscopic observing modes from $\sim$ 5 to 28~\micron. The imaging mode is acquired with the so-called MIRIM instrument (\citep{Bouchet2015}). 
In this paper, we will for the first time explore the iconic SN~1987A with investigating the heterogeneity of its temperature, morphology, and composition.
In \S\ref{sec:obs} we describe the observational program, and in \S\ref{sec:results} we present the results of the MIRIM data, including maps of the temperature variations.
We discuss the findings of this paper in \S\ref{sec:discussion}, including a description of efforts to locate the compact object left behind from the explosion (\S\ref{sec:compact}), as well as examinations of emission outside the ER and that of the outer rings (\S\ref{sec:outside the ER}).
Section \ref{sec:conclusions} presents our summary and conclusions.
Finally, details of the reduction method are described in Appendix \ref{sec:Appendix}.

\section{Observations} % Section 2
\label{sec:obs}

SN~1987A was observed with MIRI on 16 July 2022, corresponding to day 12,927 after outburst, as part of guaranteed time program \#1232 (PI:\ G.~Wright). 
Due to the brightness of the target, the subarray BRIGHTSKY, with 512$\times$512 pixels, was chosen instead of the 1024$\times$1024 pixel full field of view of MIRI, which has a detector plate scale of 0.11~arcsec pixel$^{-1}$.

As the ER has a diameter of slightly over two arcsec ($\sim$20 pixels), the size of our region of interest is well within the dimensions of this subarray.
Note as well that utilizing BRIGHTSKY lowers the sampling time from 2.775 seconds to 0.865 seconds, allowing for longer ramps for our observations.
%so longer ramps could be chosen.
We obtained images taken with four different filters:
F560W, F1000W, F1800W, and F2550W, which were chosen to sample MIRI's full 5-28~$\mu$m wavelength coverage.

\begin{figure}[t] % Figure 1
  \includegraphics[width=8.5cm]{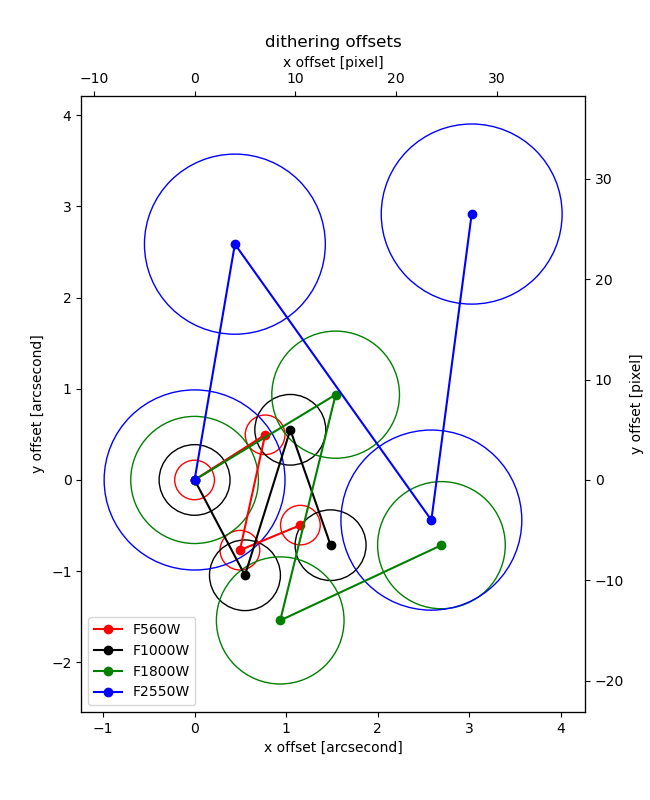}
  \caption{Illustration of the dithered positions for each of the four filters, with the Airy disk corresponding to each wavelength shown. The radius of the discs is 1.22 lambda/D, with D = 6.50~m (reminder: this is the radius of the first black ring). }
  \label{fig:dithering}
\end{figure}

%\begin{figure*}[t]
  %\includegraphics[width=9cm]{plot_blues_logz.png}
  %\includegraphics[width=9cm]{plot_greens_logz.png}
  %\includegraphics[width=9cm]{plot_reds_logz.png}
  %\includegraphics[width=9cm]{plot_3colors_log.png}
  %\includegraphics[width=18cm]{Figure2_revue.jpg}
  %\caption{MIRI images of the full 512x512 BRIGHTSKY sub-array, with the F560W (top-left), F1000W (top-right), and F1800W %(bottom-left). The F2550W image is not shown. The bottom-right panel shows a false-color image produced from the three %filters with RGB = F1800W, F1000W, F560W.}
  %\label{fig:full_field_figure}
%\end{figure*}

%\begin{figure*}[t]
 % \includegraphics[width=9cm]{plot_blues_log.png}
  %\includegraphics[width=9cm]{plot_greens_log.png}
  %\includegraphics[width=9cm]{plot_reds_log.png}
  %\includegraphics[width=9cm]{plot_3colors_logy.png}
  %\caption{MIRI images of the full 512x512 BRIGHTSKY sub-array, with the F560W (top-left), F100W (top-right), and F1800W (bottom-left). The F2550W %image is not shown. The bottom-right panel shows a false-color image produced from the three filters with RGB = F1800W, F1000W, F560W.}
  %\label{fig:full_field_figure}
%\end{figure*}

\begin{figure*}[!th] % Figure 2
\includegraphics[width=\textwidth]{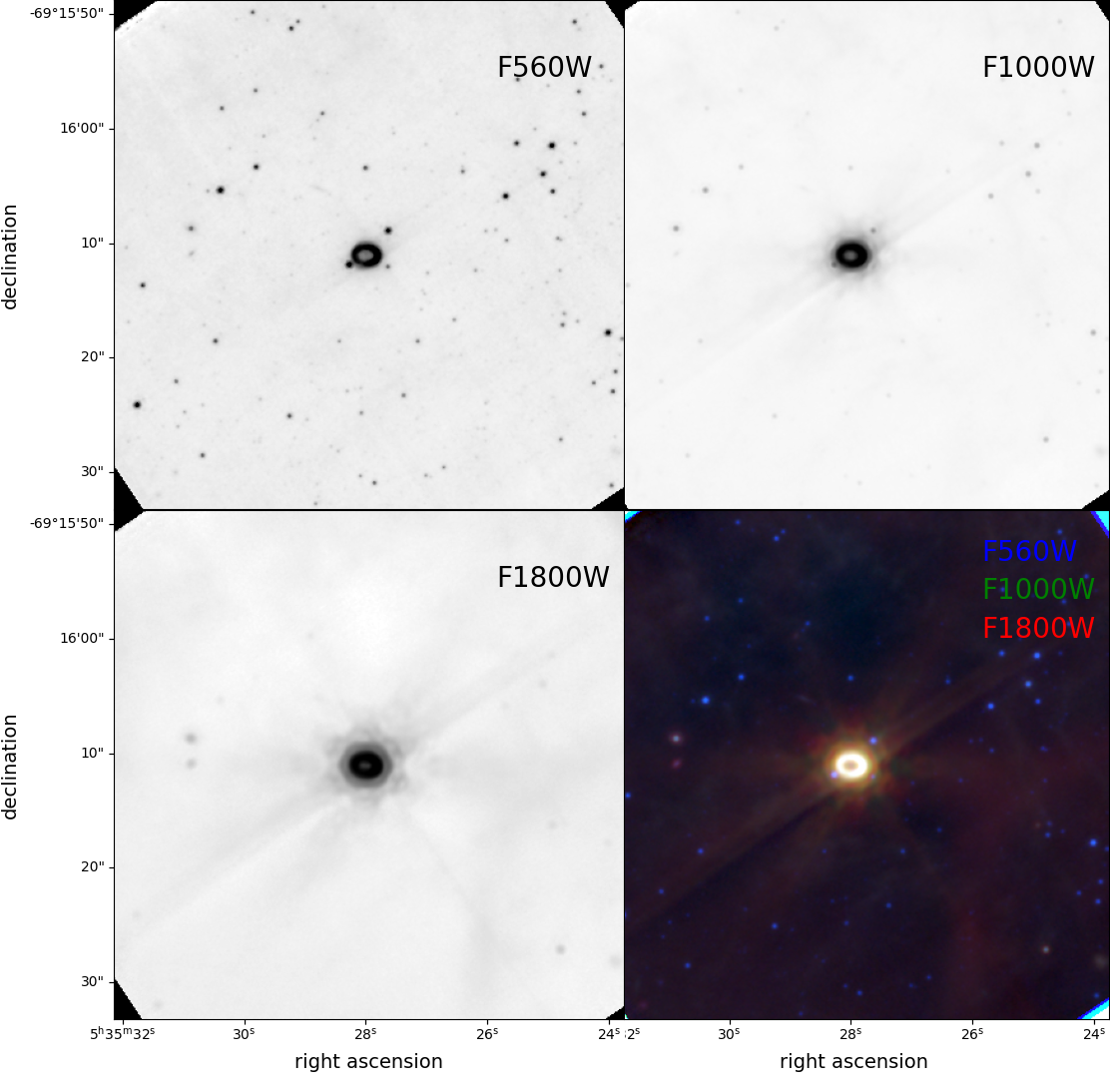}
\caption{MIRI images of the full $512\times512$ BRIGHTSKY sub-array.Images with the F560W, F1000W, and F1800W filters are shown in the top row and in the bottom-left panel. The F2550W image is not shown. The bottom-right panel shows a false-colour image produced from the three filters with RGB = F1800W, F1000W, F560W.}
  \label{fig:full_field_figure}
\end{figure*}

\begin{figure*}[th] % Figure 3
  \includegraphics[width=9cm]{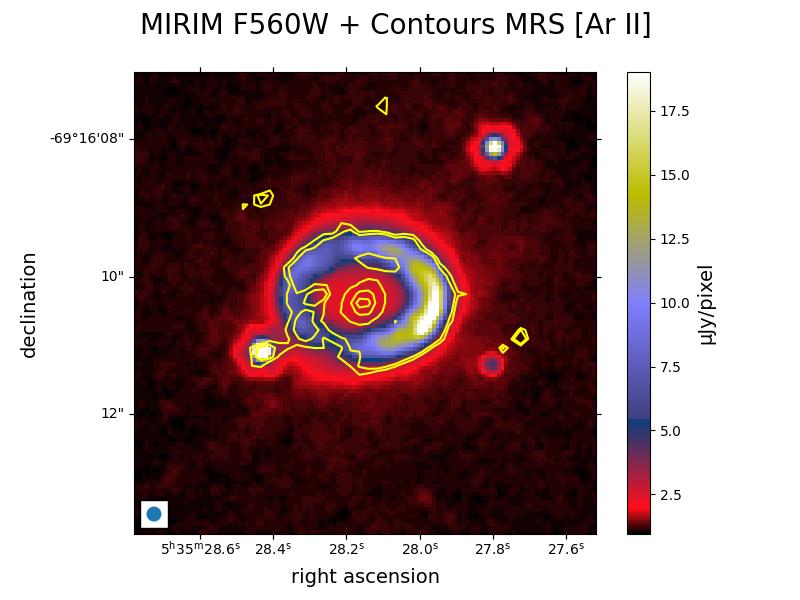}
  \includegraphics[width=9cm]{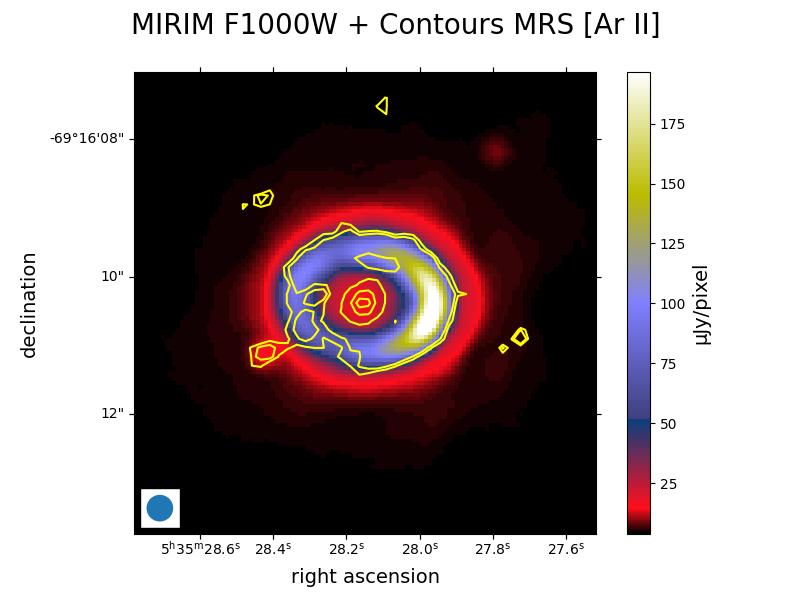}
  \includegraphics[width=9cm]{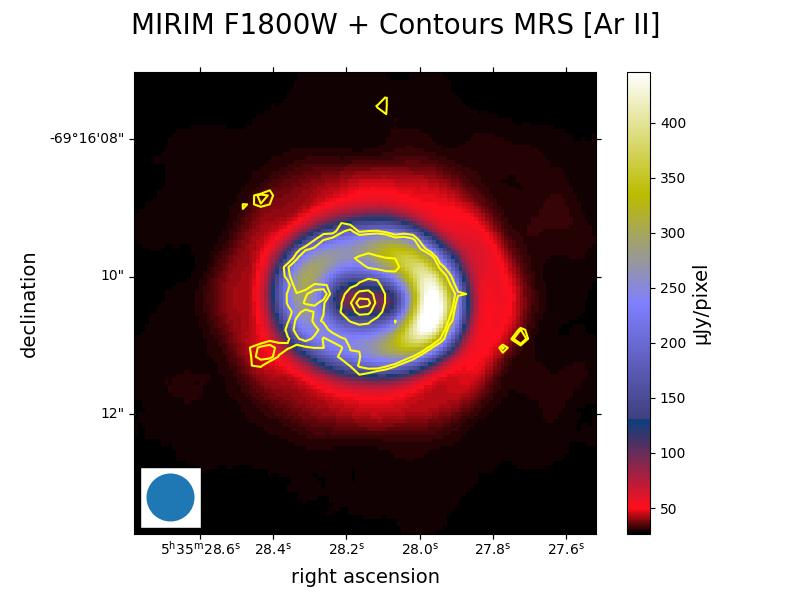}
  \includegraphics[width=9cm]{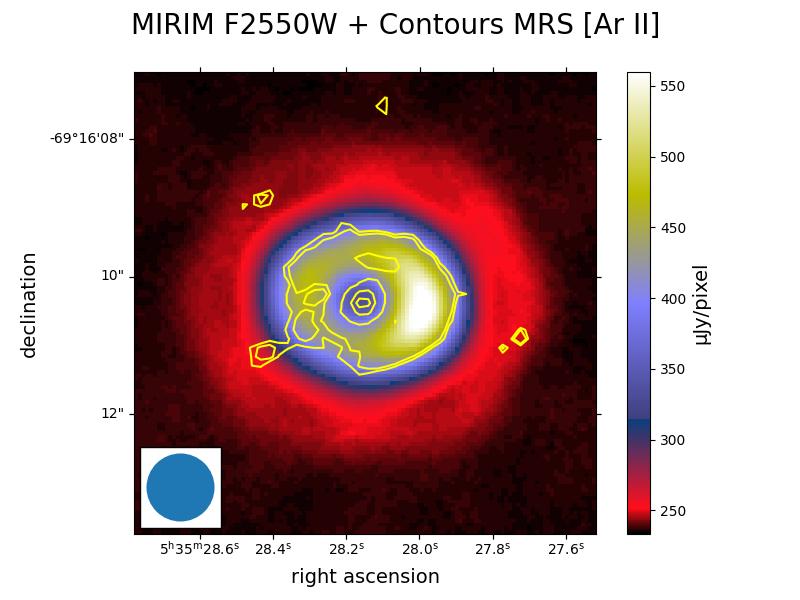}
  \caption{Final processed images at 5.6, 10, 18 \& 25.5~\micron{}, on which we superimpose the MRS [\ion{Ar}{2}] 6.985-$\mu$m contours in white. Contours levels: [90, 100, 200, 600, 1000 MJy sr$^{-1}$]. The corresponding clean beams (Gaussian fit of the actual PSF) for each wavelength are shown in the lower left corners.}
  \label{fig:4images}
\end{figure*}

\begin{figure*}[t] % Figure 4
  \includegraphics[width=18cm]{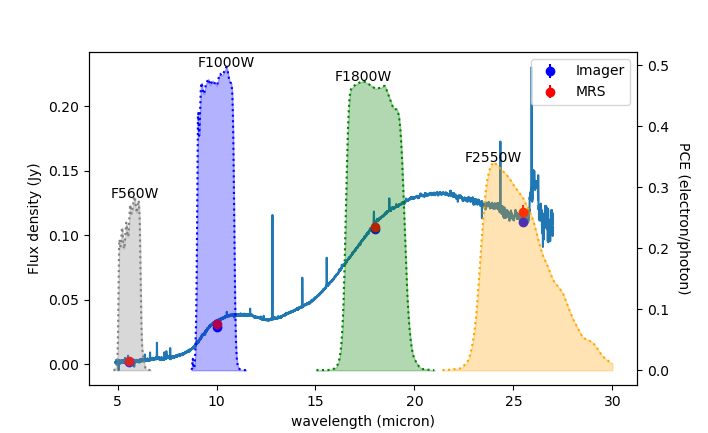}
  \caption{The `total' MRS spectrum of SN~1987A is shown, together with the photon conversion efficiency (PCE) profiles for the four filters used in MIRI imaging observations.
  The actual PCE of the F2550W filter extends to $\sim30~\mu$m, beyond the plotted MRS calibration limit.
  The blue filled circles correspond to the measured MIRIM in-band photometric fluxes, while the red filled circles correspond to equivalent in-band fluxes obtained by convolving the MRS spectrophotometry with the filter PCE profiles.}
  \label{fig:mrs_sed_filters}
\end{figure*}

The length of each integration (number of frames taken) was 16 for F560W and 15 for the other filters where fluxes are higher; these lengths were chosen so that the bright pixels would not saturate.
The number of integrations was 18 for F560W, 17 for F1000W and F1800W, and 20 for F2550W (see \citealp{Ressler2015PASP} for an explanation of the MIRI readout sensor chip assembly). For each filter, we used the standard MIRI dither pattern with four positions as shown in  Figure~\ref{fig:dithering}, which also displays the Airy disk for each filter.
Since this was one of the earliest uses of MIRI's subarray observation mode BRIGHTSKY, we describe in the Appendix the steps that were taken to reduce the data.

\section{Results} % Section 3
\label{sec:results}

\subsection{Images} % Section 3.1

Our data treatment produces the images shown in Figure~\ref{fig:full_field_figure}, which shows the BRIGHTSKY mode's 512$\times$512-pixel field of view in each of the F560W, F1000W, and F1800W filters, together with a false-colour image produced from those three filters.
The images show extended nebulosity, mostly towards the edges of the field, around a cavity with an angular diameter of $\sim$30~arcseconds that appears to surround SN~1987A. 

Figure~\ref{fig:4images} shows the central 65$\times$65 pixels in each of the F560W, F1000W, F1800W, and F2550W filters.
For each filter we have superimposed the MIRI-MRS image contours of the strong [\ion{Ar}{2}] 6.985-$\mu$m line \citep{Fransson2023}, which were derived from the data presented in \citet{Jones2023}.
Table~\ref{tab:fluxes} lists the in-band photometric fluxes measured for each of the four filters for SN~1987A's ER and ejecta.
These were measured using the same on-source and background apertures as defined by \cite{Jones2023} for their `total' MRS spectrum of SN~1987A.
An updated version of the `total' MRS spectrum is shown in Figure~\ref{fig:mrs_sed_filters}.
The raw data have been reprocessed using updated calibration reference files which primarily affect the shape of the spectrum at longer wavelengths.
The reprocessed data and spectrum will be presented in a future work.
The uncertainty in the MIRIM absolute flux calibration is estimated to be around 3-5 per~cent (Gordon et al, in prep.).
The equivalent filter fluxes obtained by convolving the MRS spectrophotometry with the photon conversion efficiency profiles for each filter plotted in Figure~\ref{fig:mrs_sed_filters} are also listed in the table. {\bf} In the case of the F2550W filter, the MRS spectrophotometric calibration cuts off at 27~$\mu$m, below the upper limit of the F2550W filter at $\sim30~\mu$m. We therefore extrapolated the continuum above 27~$\mu$m using a second order polynomial fit to the continuum in the 23-27~$\mu$m range to account for `missing’ flux in the MRS spectrum that would be detected by MIRIM.
The uncertainty in the MRS absolute spectrophotometric flux calibration is estimated by \cite{Argyriou2023} to vary from 4~per~cent at its shorter wavelengths (Bands 1 and 2) to 5~per~cent (Band 3) and 6~per~cent (Band 4) at its longer wavelengths.
The MIRIM and equivalent MRS photometric fluxes are plotted as the blue and red solid symbols in that figure.
Inspection of Table~\ref{tab:fluxes} shows excellent agreement between the Imager and equivalent MRS filter fluxes in the cases of the F560W and F1000W filters, and good agreement for the F1800W and F2550W filters.The F2550W band includes the strong [\ion{Fe}{2}] 25.99~$\mu$m + [\ion{O}{4}] 25.89~$\mu$m feature (see Figure~\ref{fig:mrs_sed_filters}) but, as listed in the last row of Table~\ref{tab:fluxes}, we find that these and other emission lines contribute only 2.9~per~cent of the total in-band flux in the F2550W filter. {\bf} We estimated the contributions of the emission lines in a given filter by comparing the calculated in-band flux of the spectrum with and without these lines. For the latter, we masked the emission lines and interpolated the continuum across the resulting gaps using a second order polynomial. 
The F560W filter has a much weaker continuum, but even for that filter emission lines are found to contribute only 1.9~per~cent of the total in-band flux.
For the F1000W and F1800W filters, emission lines make a negligible contribution (Table~\ref{tab:fluxes}).
The contributions of synchrotron and free-free emission are also small except perhaps at wavelengths shorter than 8 \micron\ \citep{Jones2023}.
It would seem that the most likely explanation for the small discrepancy at the two longest wavelengths between the Imager filter fluxes and the equivalent MRS filter fluxes is a difference in the background estimation, given that background flux levels rise steeply to longer wavelengths. 

\cite{Larsson2011}, \cite{Helder2013}, and \cite{Frank2016} reported on the brighter optical and X-ray emission in the western regions of the ring.
As already seen at optical wavelengths \citep{Fransson2015}, comparison of the day 6526 11.7 and 18.3~$\mu$m images of \cite{Bouchet2006} with the new JWST images taken at day 12927 shows that the emission from the NE region of the ring has fallen dramatically and the strongest emission from the ring now arises from the SW region.
This seems to indicate that most of the dust which was lying in the NE region has been either destroyed or has recently cooled after the passage of the shock.
The change in brightness alone does not permit to distinguish between the two scenarios.

\begin{figure}[th] % Figure 5
  \includegraphics[width=9.cm]{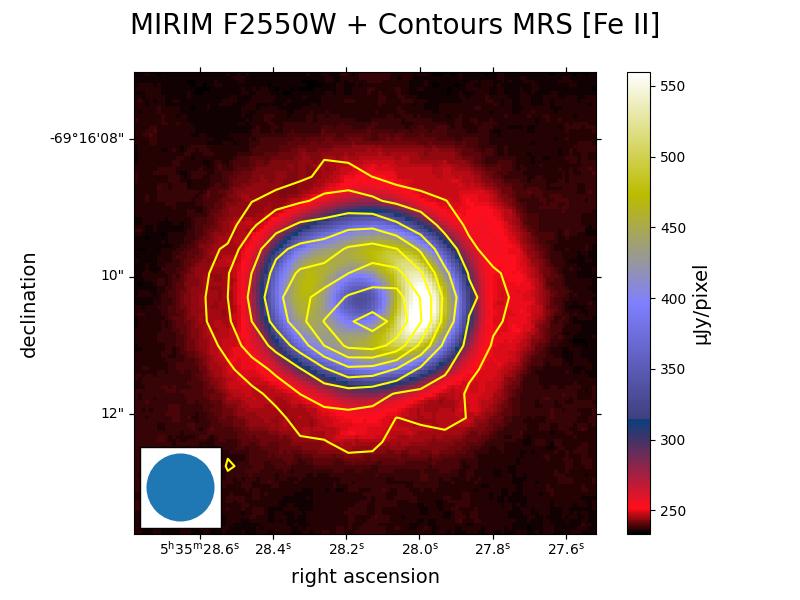}
  \caption{The MIRIM image in the F2550W filter, with the white contours showing the distribution of the [\ion{Fe}{2}] 25.99-$\mu$m + [\ion{O}{4}] 25.89-$\mu$m emission from the MRS data. Contours levels: [900, 1000, 1200, 1400, 1600, 1800, 2000, 2200 MJy/sr]}
  \label{fig:FeII_MRS}
\end{figure}

In Figure \ref{fig:FeII_MRS} we have superimposed on our image obtained through the F2550W filter the contours of the [\ion{Fe}{2}]$\lambda $25.99~\micron\ line from the MRS data.
The MRS data clearly showed that this prominent line is coming mostly from the inner ejecta and the ejecta close to the reverse shock \citep{Jones2023}, while the dust which dominates the F2550W filter is from the ring.
%It has to be noted that this line is blended with the [\ion{O}{4}]$\lambda\lambda$25.89~\micron\ line whose origin is not clear yet (See Figure \ref{fig:MRS_19-27}).
The Imager data, which have a spatial resolution $\sim$2 times better than the MRS data, confirm this conclusion, with the dust-dominated F2550W image showing a drop in emission inside the ER, whereas the [\ion{Fe}{2}]$\lambda 25.99$-dominated contours peak inside the ER.

\begin{deluxetable}{ccccc} % Table 1
\label{tab:fluxes}
\tabletypesize{\small}
\tablewidth{0pt}
\tablecaption{Filter fluxes for SN~1987A from the MIRIM and equivalent fluxes from the MRS spectrophotometry (see text).}
\tablehead{
\colhead{Flus (mJy)}&\colhead{F560W}&\colhead{F1000W}&\colhead{F1800W}&\colhead{F2550W}
}
\startdata
Imager & 1.8 $\pm$ 0.1 & 29.3 $\pm$ 0.1 & 105.9 $\pm$ 0.3 & 110.2 $\pm$ 1.4 \\ 
MRS  & 2.4 $\pm$ 0.6 & 31.1 $\pm$ 0.3 & 106.2 $\pm$ 0.5 & 118.3 $\pm$ 5.0 \\
Line & 1.9\% & $<$0.1\% & $<$0.1\% & 2.9\% \\
 Contribution & & & &
\enddata
\end{deluxetable}

\subsection{Temperature and mass maps} % Section 3.2
Several dust composition models have been proposed for SN~1987A.
After the far-IR Herschel observations revealed a large reservoir of cold dust, \cite{Matsuura2015} proposed a mix of amorphous carbon and silicates which yields to T$_{Dust}\sim23-27~K$, and M$_{Dust}\sim0.3-0.8~M_{\sun}$, depending on whether the composition arises solely from amorphous carbon or a mix with silicates.

\cite{Cigan2019} comment on the ALMA images, and conclude that the sub-mm emission coming from the inner region is due to thermal emission from the ejecta which was first pointed out by \cite{Indebetouw2014}, and then \cite{Cendes2018} included synchrotron emission.

Computing a temperature map based on a simple blackbody emission should apply for optically-thick emission from the ejecta only.
Moreover, fitting a blackbody model to the mid-IR observations is not justified since the mid-IR fluxes are dominated by optically-thin emission from silicate dust grains in the ER \citep{Arendt2016, Jones2023}.

For an optically thin point source, the flux density, $F_\nu(\lambda)$ at wavelength $\lambda$ is given by \citep{Hildebrand1983,Bouchet2006}

$$F_\nu(\lambda) = 4M_d  \frac{\kappa(\lambda)\pi B_\nu(\lambda, T_d)}{4\pi D^2} , $$
where $M_d$ is the dust mass, $\kappa(\lambda)$ is the dust mass absorption coefficient at wavelength $\lambda$, $B_\nu(\lambda,T)$ is the Planck function, and $D$ is the distance to the supernova, taken to be $D = 51.4$ kpc \citep{Panagia1991}.

\begin{figure}[t] % Figure 6
  \includegraphics[width=9.cm]{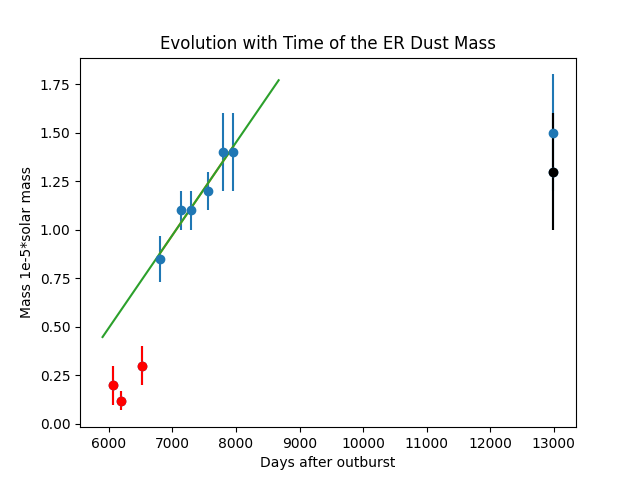}
  \caption{Evolution with time of the ER dust mass (red dots are from \cite{Bouchet2006}; blue dots are from \cite{Jones2023}; black dot is from the present work. The linear increase during the period [6805, 7955] day is fitted by :
$M_\sun = 4.77~10^{-4} \times t_{Day} - 2.37$)}
  \label{fig:massevolution}
\end{figure}

\begin{figure}[t] % Figure 7
  \includegraphics[width=9.cm]{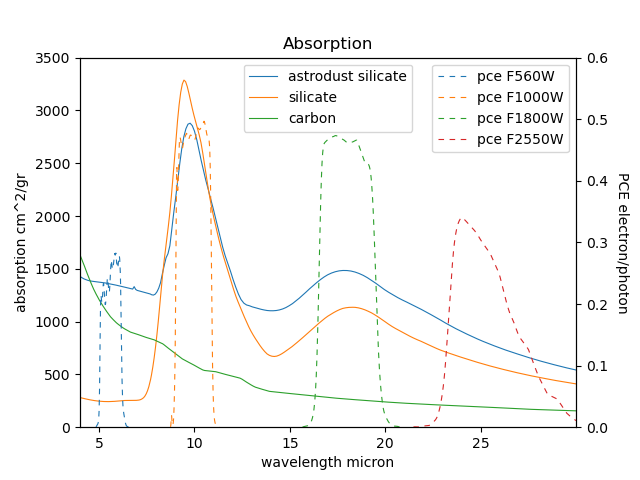}
  \caption{Dust mass absorption coefficients for some of the grain species (references in the text) and the Photon-to-electron Conversion Efficiency (pce) for each filter.}
  \label{fig:abscoeff}
\end{figure}

\begin{figure*}[t] % Figure 8
\includegraphics[width=9cm]{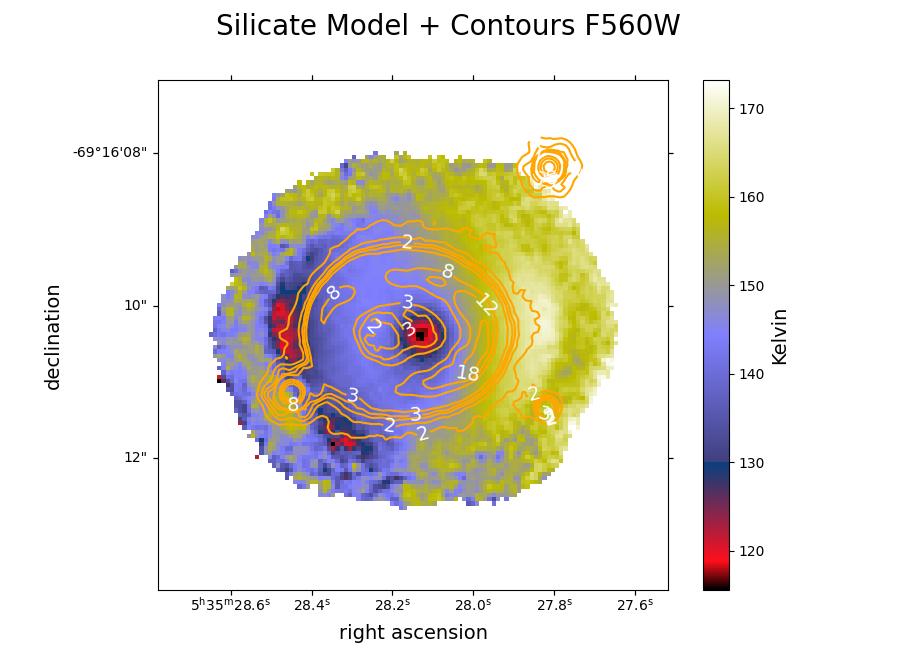}
\includegraphics[width=9cm]{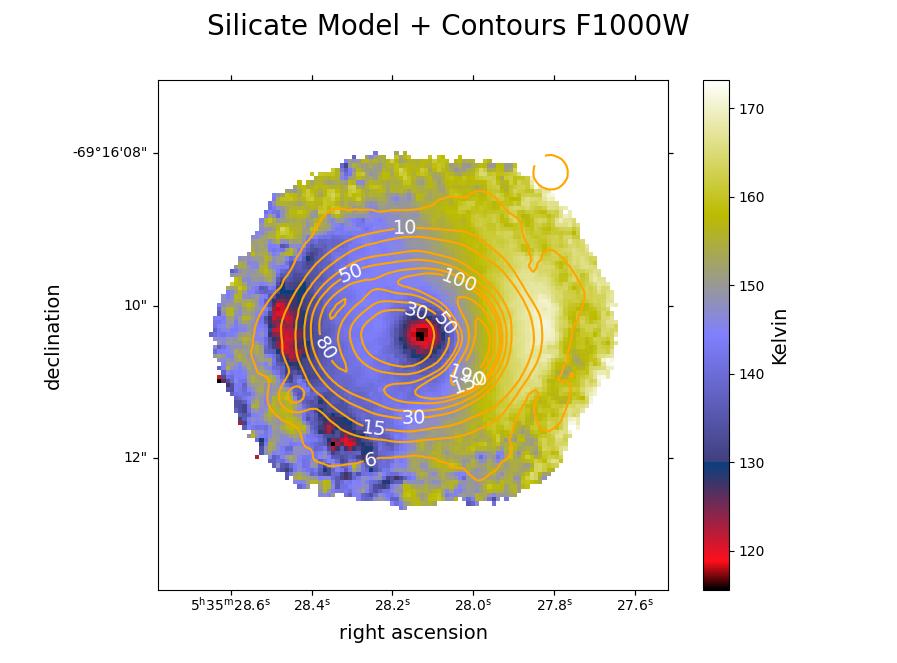}
\includegraphics[width=9cm]{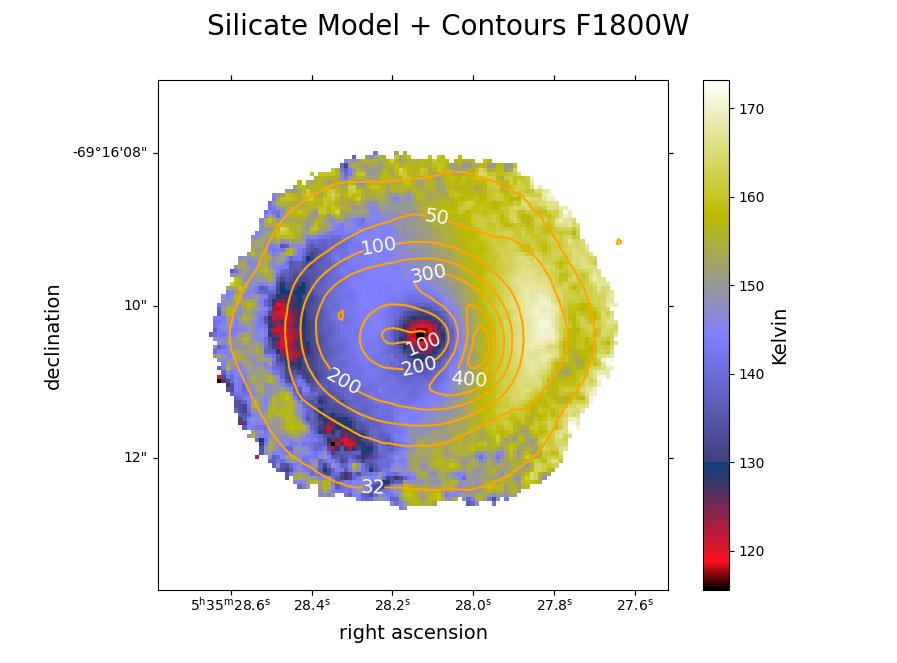}
\includegraphics[width=9cm]{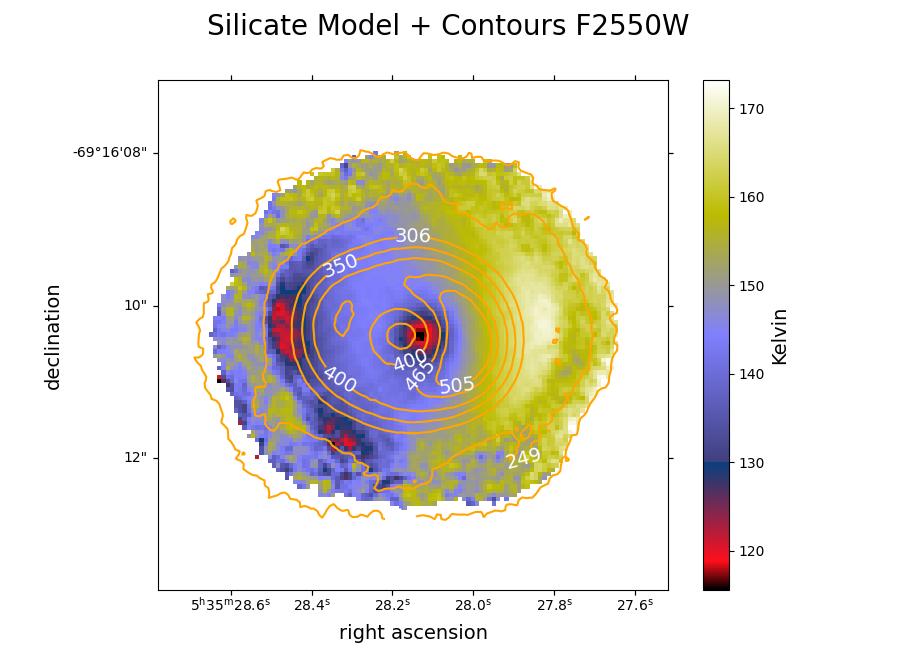}
\caption{Temperature map computed with the silicates and amorphous carbon model as discussed in the text, with contours from the 4 images. Contour levels: F560W: [1.5, 2.2, 2.4, 2.7, 3.2, 8, 12, 18 $\mu$Jy per pixel]; F1000W: [6, 10, 15, 30, 50, 80, 100, 150, 190 $\mu$Jy per pixel]; F1800W: [35,50,100, 200, 300,400, 450 $\mu$Jy per pixel]; F2550W: [240, 248.6, 306, 350, 400, 465, 505 $\mu$Jy per pixel]}
\label{fig:sili}
\end{figure*}

We first fitted the integrated MIRI flux densities with a population of dust particles consisting of a single population of silicate and amorphous carbon grains to account for the shorter wavelengths emission as explained in \cite{Bouchet2006} and \cite{Arendt2016}. We used carbon and silicate dust mass absorption coefficients from \citet[][see Figure~\ref{fig:abscoeff}]{Draine1984}, with results shown in Figure \ref{fig:sili} and Figure~\ref{fig:masses} (the four filters have been used to generate these maps). 
The temperature computed from this model varies from 120~K to 165~K and the mass varies from $0.2$ to $1.5~10^{-8}M_\sun$ per pixel. The total mass, that is the sum of the pixels enclosed by the contours (equatorial ring) is then $1.3\pm0.5~10^{-5}M_\sun$, {\bf} similar to the mass reported by \cite{Jones2023} for the same day ($1.5\pm0.3~10^{-5}M_\sun$). It is striking to note that although the temperature is similar, the total mass of the dust is $\sim$10 times greater than those reported by \cite{Bouchet2006}. It should be emphasized that the essential difference between these two dates is the amount of material that interacts with the ER. \cite{Bouchet2006} report masses of $0.1 - 0.2\pm0.03~10^{-5}M_\sun~$ at days 6070 and 6190, and $0.3\pm0.1~10^{-5}M_\sun~$ at day 6525, which marks about the onset of the interaction of the ejecta with the equatorial ring and the dust mass in the ER. \cite{Jones2023} report that this mass was already $0.85\pm0.15~10^{-5}M_\sun$ at day 6805, then increased linearly until day 7955 (during that period this linear increase is fitted by:  
$M_\sun = 4.77~10^{-4} \times t_{Day} - 2.37$),
to remain constant up to day 12,927 (Table~\ref{tab:mass_evolution} and Figure~\ref{fig:massevolution} summarise the dust mass evolution). That means that the interaction of the ejecta with the ER was accompanied by an increase in dust mass.  As a result, we have two possible explanations: (i) either this interaction makes a new condensation of grains, (ii) or a fraction of the very cold dust initially in the ejecta has warmed up in the ring environment and therefore the mass of warm dust increased while that of very cold dust detected by {\it Herschel} and ALMA would have decreased  (future ALMA observations should confirm this hypothesis). 

\begin{deluxetable}{rcc} 
\label{tab:mass_evolution}
\tabletypesize{\small}
\tablewidth{0pt}
\tablecaption{Evolution with time of the Dust Mass in the ER}
\tablehead{
\colhead{Day}&\colhead{Mass($10^{-5}~M_\sun$)  }&\colhead{Reference} }

\startdata
6070& 0.20 $\pm$ 0.10 & \cite{Bouchet2006} \\
6190& 0.12 $\pm$ 0.05 & \cite{Bouchet2006} \\
6526& 0.30 $\pm$ 0.10 & \cite{Bouchet2006} \\
6805& 0.85 $\pm$ 0.12 & \cite{Jones2023} \\
7138& 1.10 $\pm$ 0.10 & \cite{Jones2023} \\
7296& 1.10 $\pm$ 0.10 & \cite{Jones2023} \\
7555& 1.20 $\pm$ 0.10 & \cite{Jones2023} \\
7799& 1.40 $\pm$ 0.20 & \cite{Jones2023} \\
7955& 1.40 $\pm$ 0.20 & \cite{Jones2023} \\
12997& 1.50 $\pm$ 0.30 & \cite{Jones2023} \\
12997& 1.30 $\pm$ 0.30 & Present work
\enddata
\end{deluxetable}

\begin{figure*}[t] % Figure 9
\includegraphics[width=9cm]{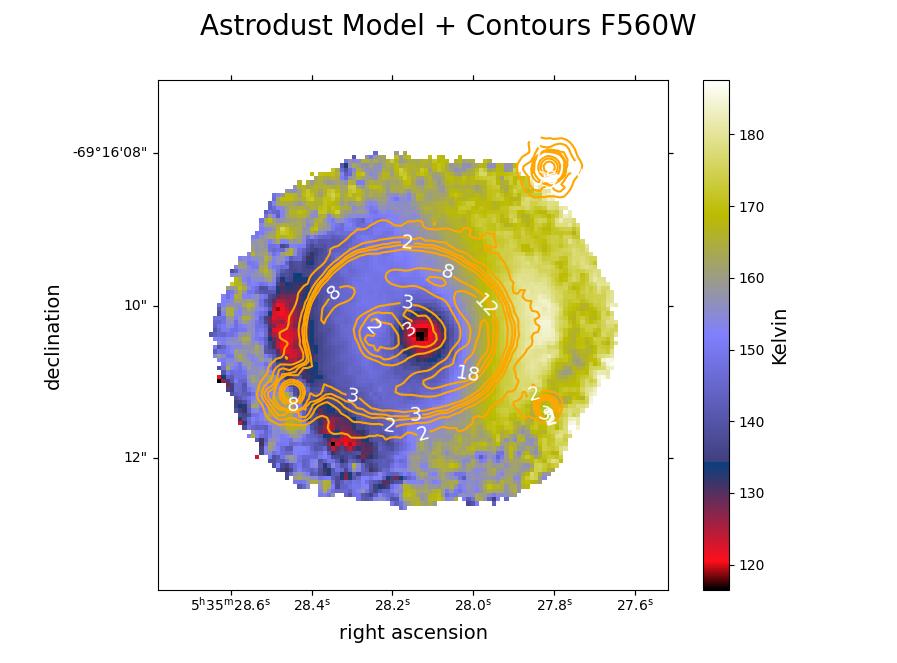}
\includegraphics[width=9cm]{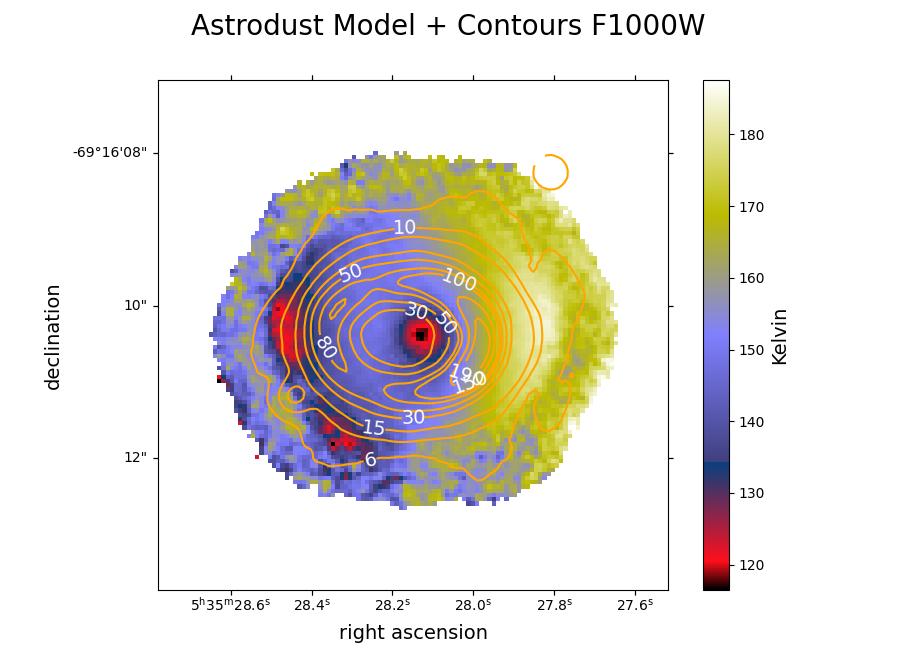}
\includegraphics[width=9cm]{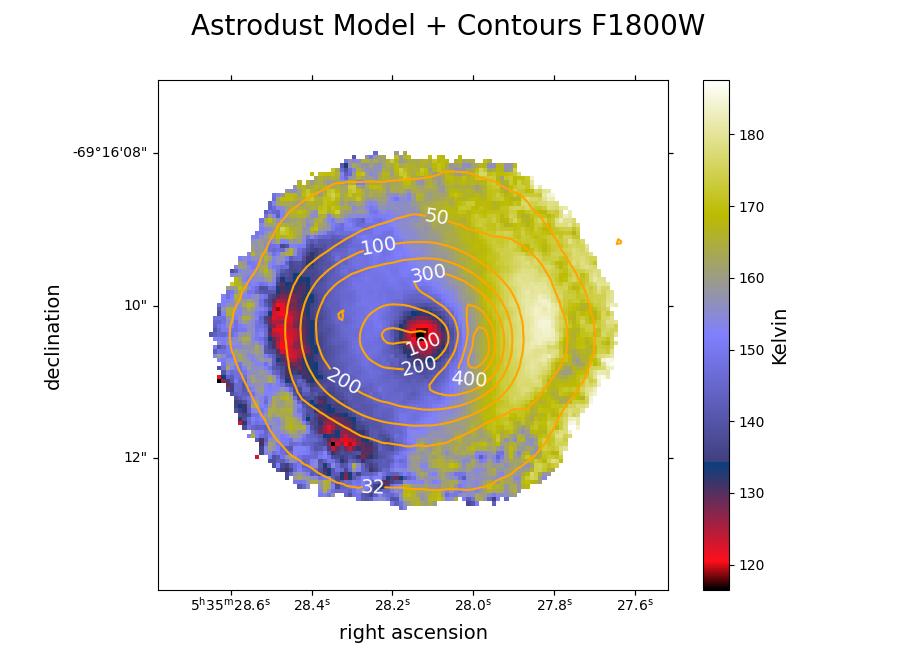}
\includegraphics[width=9cm]{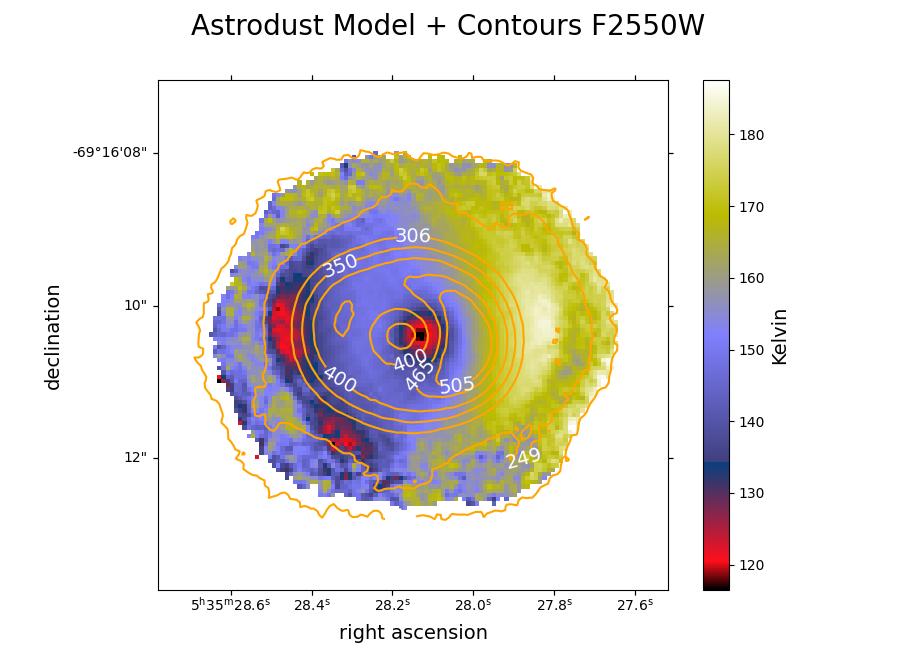}
\caption{Temperature map computed with the {\em astrodust} model with the data from the 4 filters as discussed in the text, with contours from the 4 images as in Figure 7. Contour levels are indicated on the figure.}
\label{fig:astrodust}
\end{figure*}

\begin{figure*}[t] % Figure 10
\includegraphics[width=9cm]{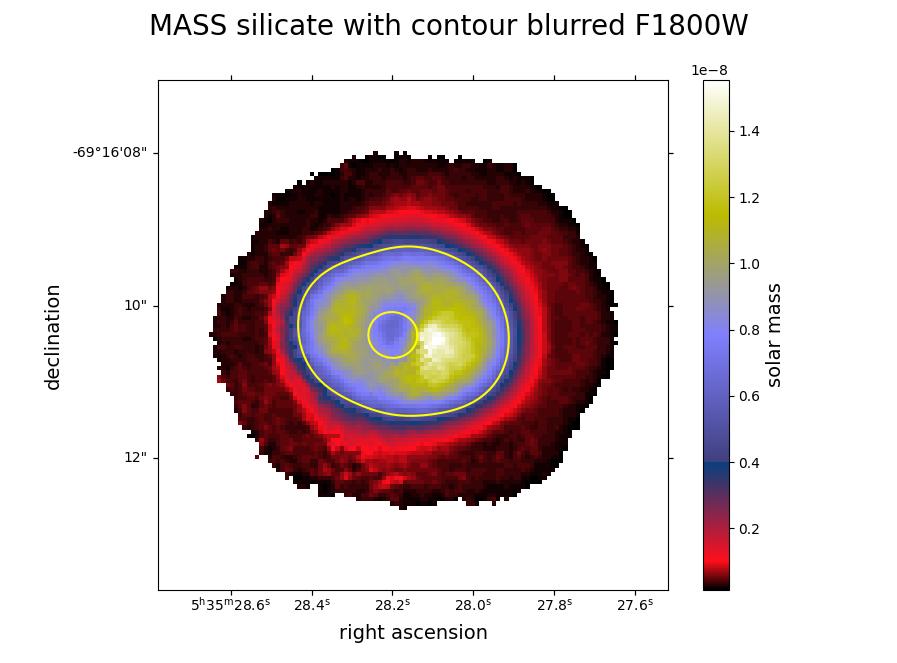}
\includegraphics[width=9cm]{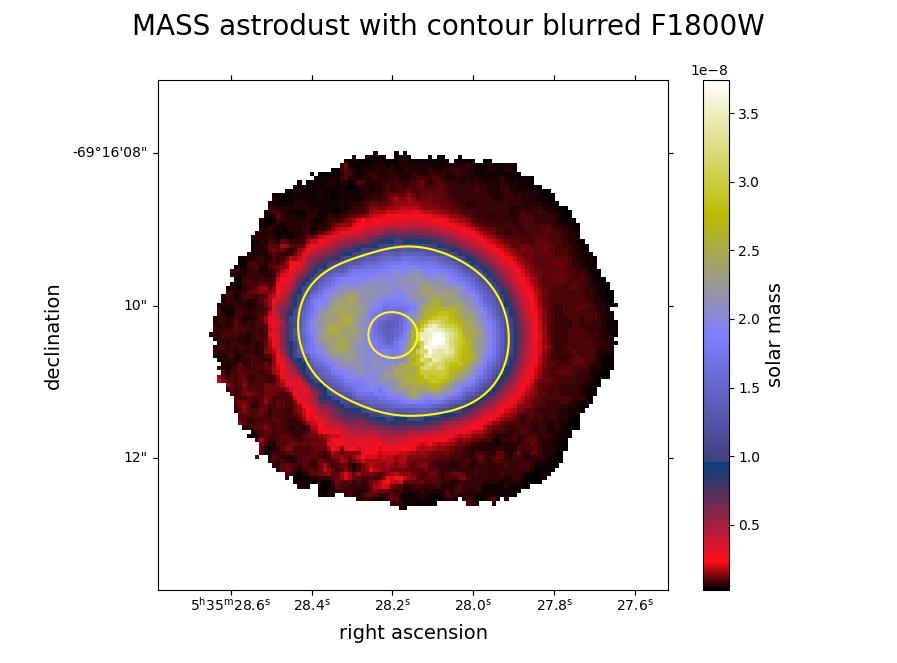}
\caption{Mass maps computed with the two dust grain compositions : silicates (left) and {\em astrodust} (right).  The contour level is 150~$\mu$Jy per pixel that delimits the equatorial ring (the total mass is the sum of the pixels enclosed by the contour). }
\label{fig:masses}
\end{figure*}

We also generated dust temperature and mass maps using the so-called {\em astrodust} model, as described by \cite{Hensley2023} and applied to SN 1987A by \cite{Jones2023}. Results are shown in Figure~\ref{fig:astrodust} and Figure~\ref{fig:masses} (as for Figure~\ref{fig:sili} the four filters have been used to generate these maps). The temperature varies from 125~K to 175~K, quite consistent although slightly warmer than the temperatures computed from the silicates model, and the mass varies from $0.5$ to $3.5~10^{-8}M_\sun$ per pixel. The total mass using the {\em astrodust} composition is then $2.8~10^{-5}M_\sun$, significantly higher that the mass reported by \cite{Jones2023} ($1.7\pm0.2~10^{-5}M_\sun$).

Note that the contours shown in Figure \ref{fig:sili} and \ref{fig:astrodust}  delineate the ER, within which the models make sense.
The composition of the dust beyond the ER is not known, and it is doubtful that the models can apply. Both models show that the warmest temperatures are located on the outer edge west of the ER. 
The comparison of the mass absorption coefficient for each of the grain species is shown in Figure \ref{fig:abscoeff}.

\begin{figure*}[t] % Figure 11
\includegraphics[width=6.5cm]{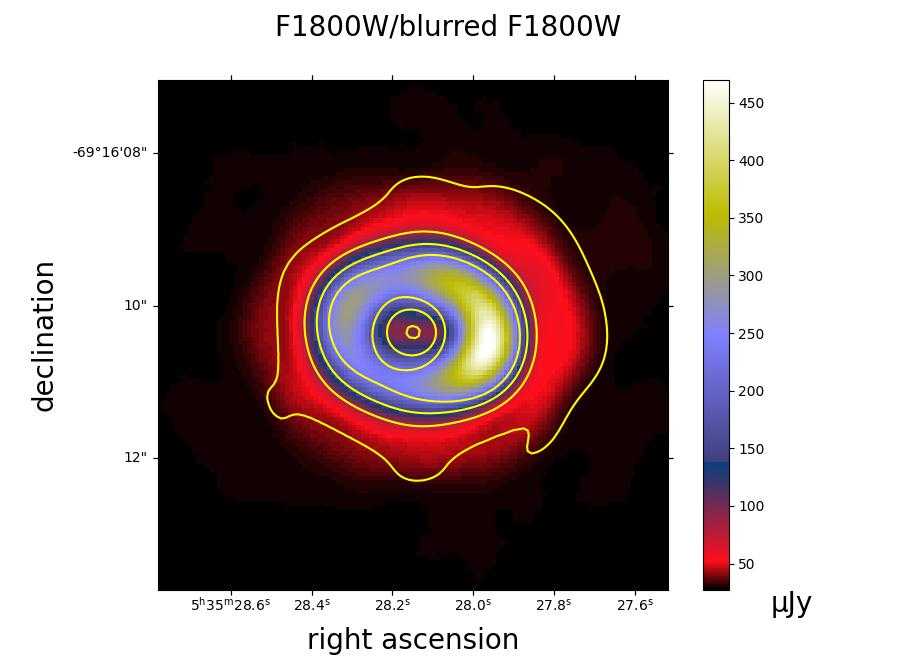}
\includegraphics[width=6.5cm]{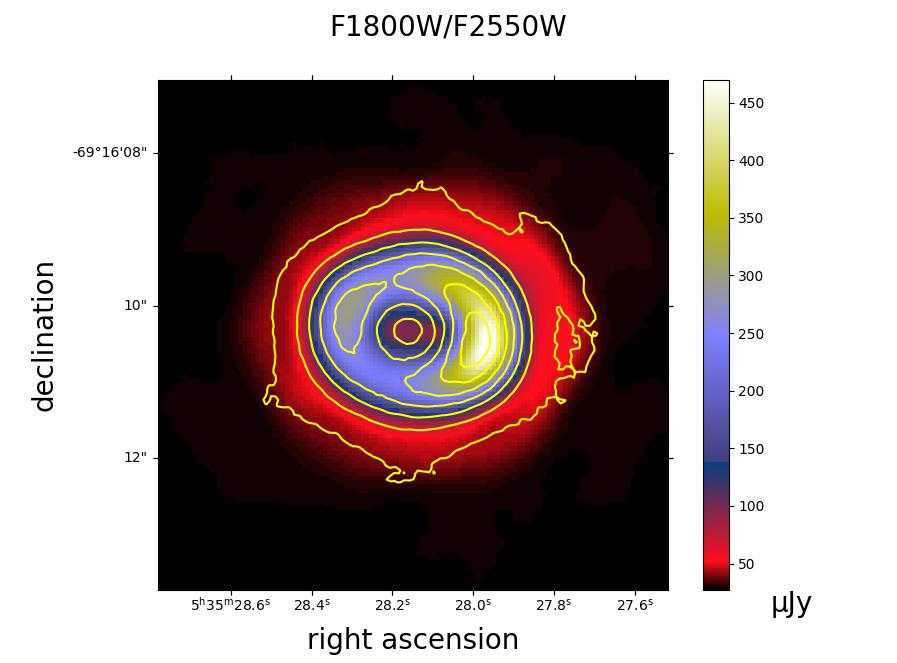}
\includegraphics[width=6.5cm]{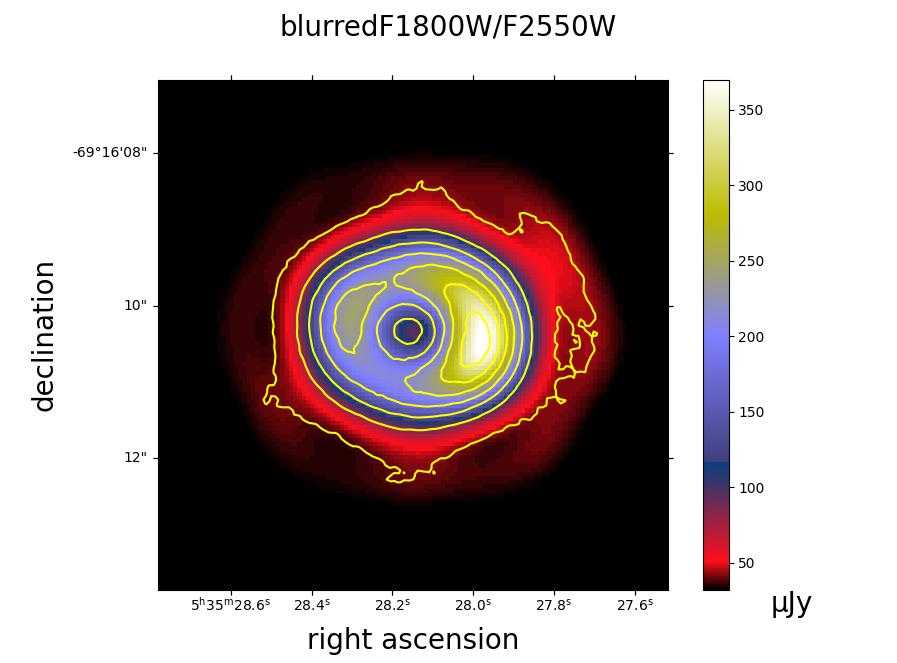}
\caption{Left: MIRI image obtained with the filter F1800W with contours from the same image blurred at the resolution of the F2550W image; middle: image obtained with the filter F1800W with contours from the F2550W image; right: image obtained with the filter F1800W blurred at the resolution of the F2550W filter and contours from the F2550W image .}
\label{fig:F18_blurred_contours25}
\end{figure*}

\begin{figure*}[t] % Figure 12
\includegraphics[width=6.5cm]{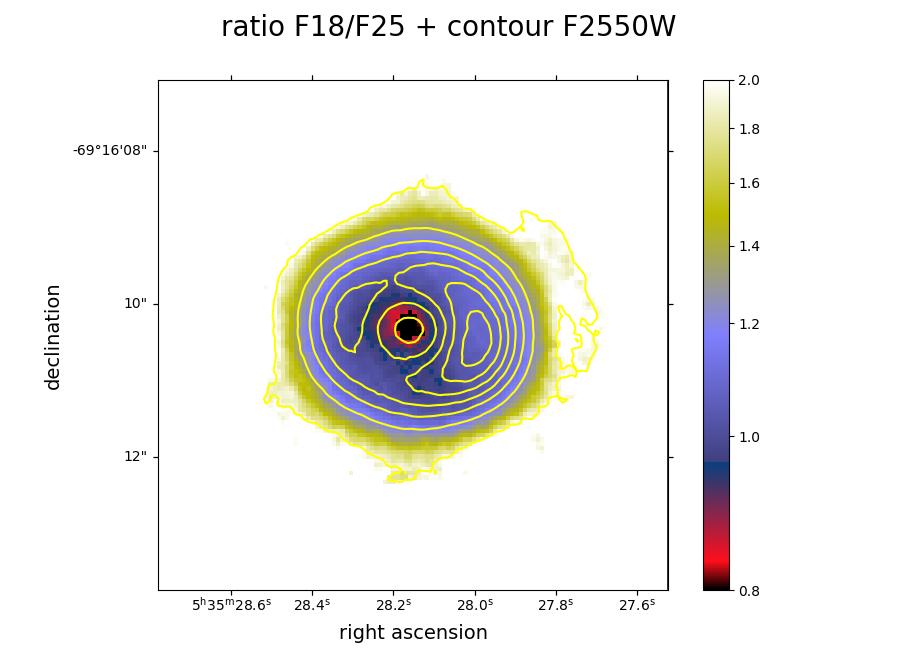}
\includegraphics[width=6.5cm]{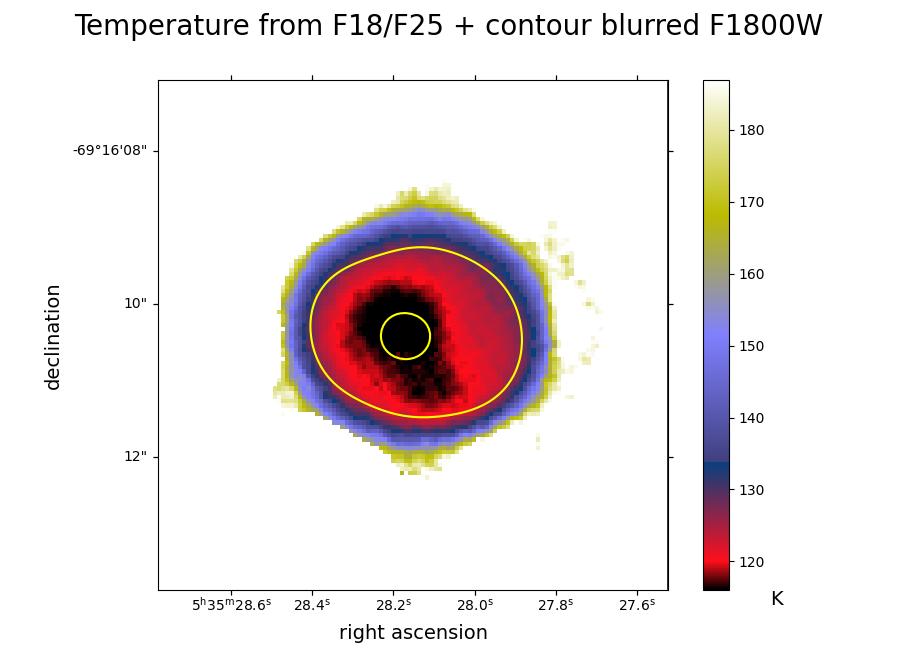}
\includegraphics[width=6.5cm]{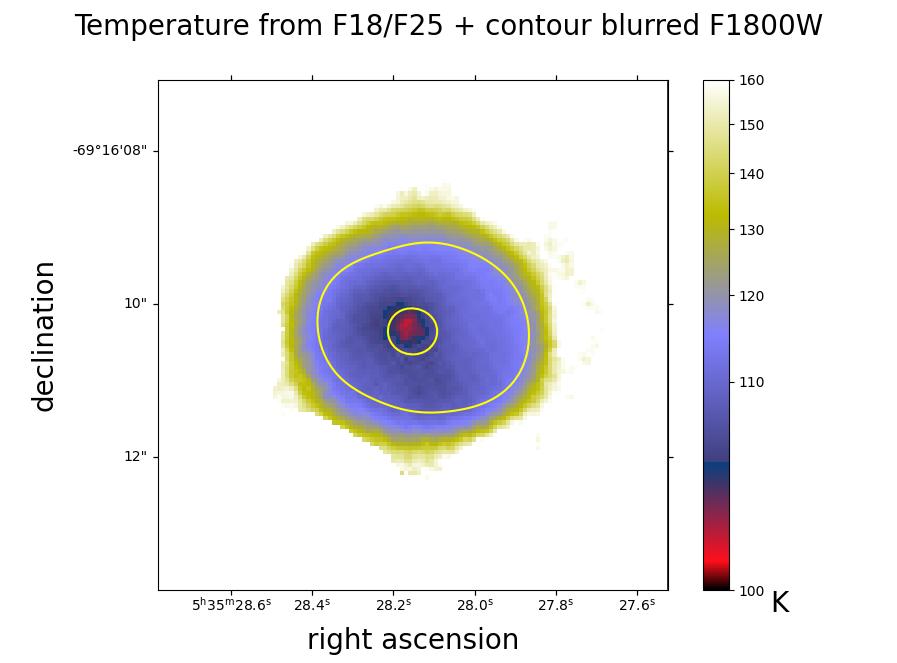}
\caption{Left: ratio of the blurred F1800W over the F2550W images, with contours from the F2550W image [250, 300, 350, 400, 450, 500, 550 $\mu$Jy per pixel]; middle: Temperature map computed with the {\em astrodust} model applied to the F1800W over the F2550W filters ratio, with contours from the blurred F1800W image. To compare with Figure~\ref{fig:astrodust}. Contours levels are 150 $\mu$Jy per pixel, which define 2 ellipses that enclose the ER; right:  same as the middle figure with a  logarithmic color scale in order to pinpoint the central region.}
\label{fig:ratio-temp}
\end{figure*}

{\bf} In order to estimate temperatures (electron, ER  and ejecta) \cite{Jones2023} use various lines ratios. According to these authors, the strongest emission line in the F1800W MIRIM filter is the [FE~II] 18.93~\micron\ line from the ER, while the F2550W filter includes the prominent [Fe~II] 26~\micron\ emission line from the ejecta. In the same vein as these authors we produce a F2550W-to-F1800W ratio image (after blurring the F1800W image to the lower resolution of the F2550W image; see Figure~\ref{fig:F18_blurred_contours25}) and we derive a temperature map using these two filters only with the {\em astrodust} model \citep{Hensley2023}.
The result is shown in Figure~\ref{fig:ratio-temp}. Although the dust temperature range of 105~K to 140~K is consistent with the previous calculation made with the four filters (Figure~\ref{fig:astrodust}), the west side of the ER appears slightly cooler in this computation. {\bf} Note that part of this result should be taken with caution for it is doubtful that we can use the {\em astrodust} model to derive temperatures in the ejecta.
Also, from \cite{Jones2023}, we need two-grain components to reproduce the emission, so it is clear that a single temperature fit across the entire SED might be too influenced by the changing mass of the small grain component.

\section{Discussion} % Section 4
\label{sec:discussion}

\subsection{Searching for the compact object} % Section 4.1
\label{sec:compact}

\begin{figure*} % Figure 13
\includegraphics[width=9cm]{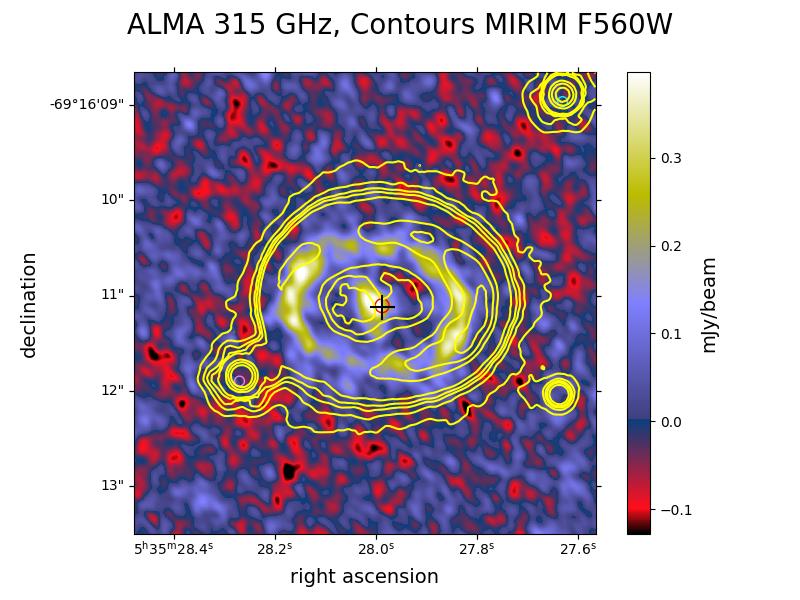}
\includegraphics[width=9cm]{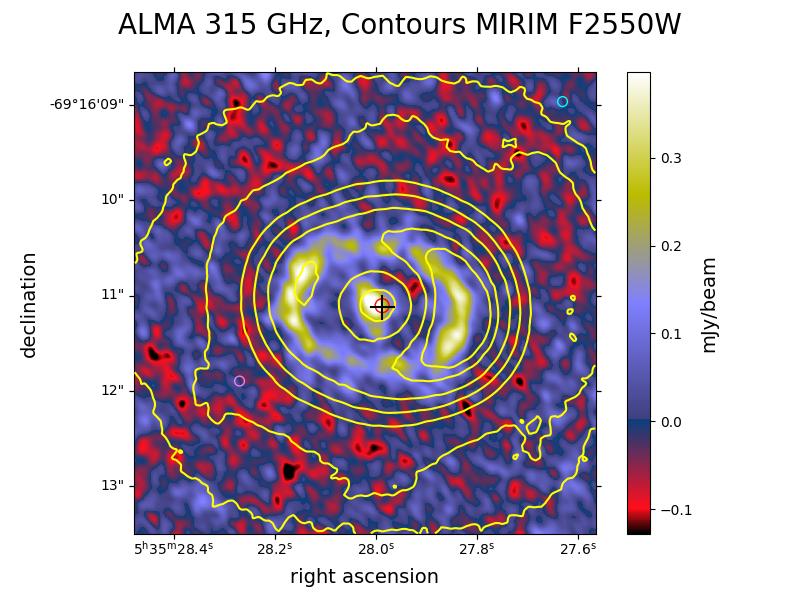}
\includegraphics[width=9cm]{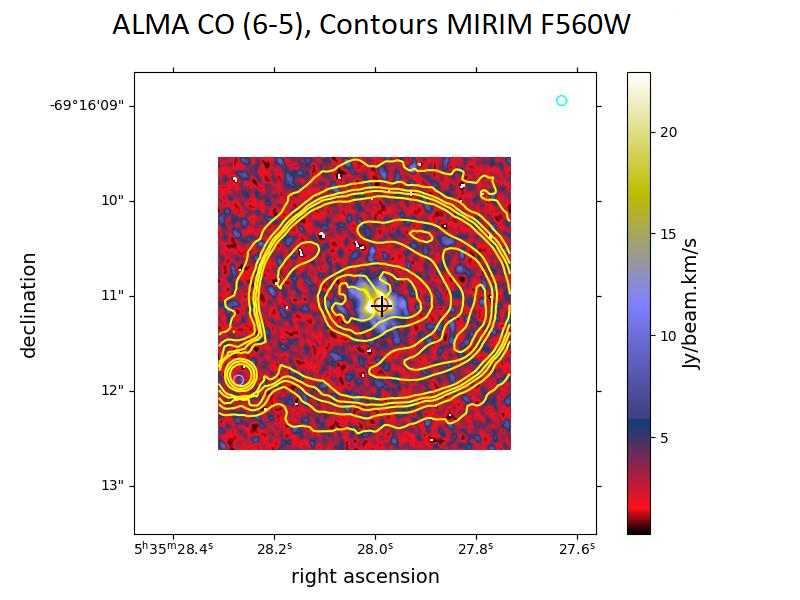}
\includegraphics[width=9cm]{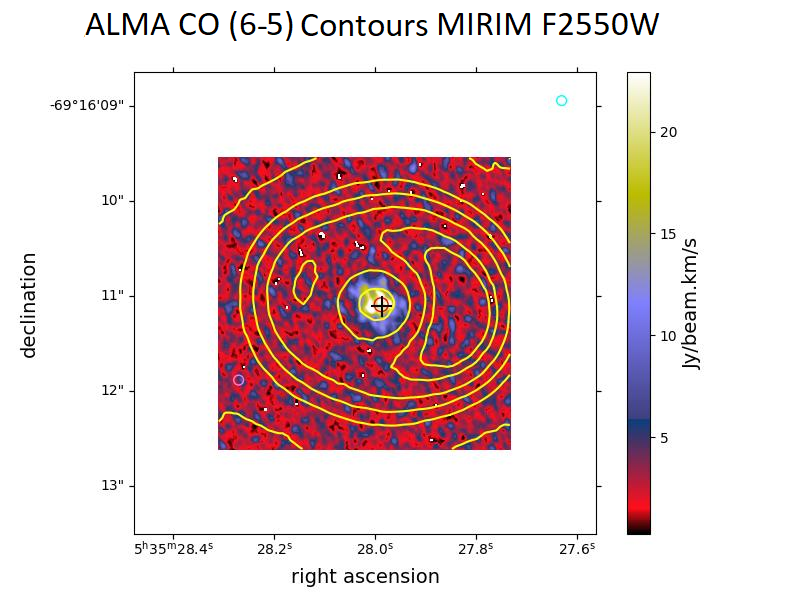}
\caption{ALMA images at 315~GHz (952~\micron) and in the CO (6-5) band at 641.47~GHz (467~\micron), contours from the MIRI 
images as indicated, with the position of the 
sub-mm dust {\em `Blob' }, proposed by \cite{Cigan2019} to be heated by a neutron star candidate denoted with a cross.}
\label{fig:alma}
\end{figure*}

\begin{figure*}[t] % Figure 14
\includegraphics[width=18cm]{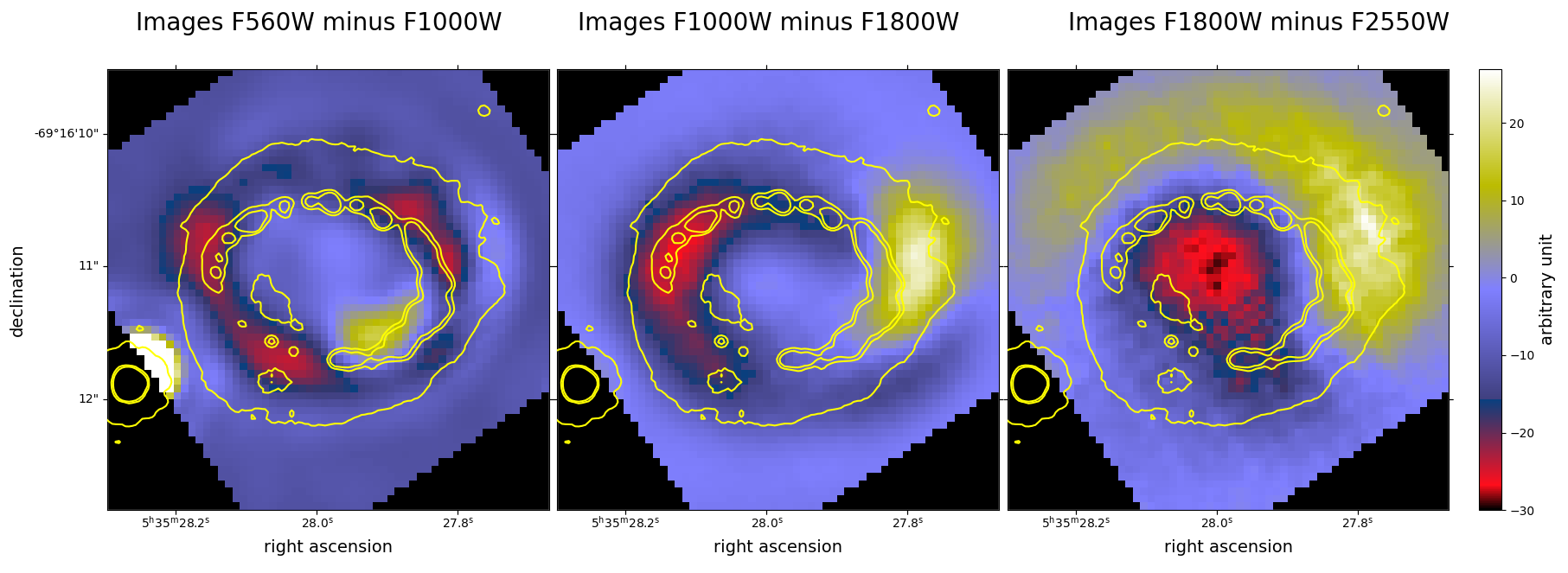}
\caption{Recursive subtraction between adjacent filters to allow comparison of the flux evolution with wavelengths. Contours from the HST/WFC3 image from 2022 (Rosu et al. in prep) are drawn to illustrate the position of the ER.
  The negative fluxes are due to the subtraction between wavelengths. Contours levels: [0.06, 0.6, 1 electrons/s]. The HST image has been offset by [22, 20] pixels.}
\label{fig:troisdiff}
\end{figure*}

The progenitor of SN~1987A, Sanduleak -69 202, was a blue supergiant \citep{West1987,White1987,Gilmozzi1987,Kirshner1987}, thought to have had a zero-age main sequence mass of $\sim$19~M\,$_\sun$ \citep{Woosley1987,Hashimoto1989AA}, with a mass of $\sim$14~M\,$_\sun$ at the time of the explosion \citep{Woosley1988,Smartt2009b,Sukhbold2016}.
From its mass, the expectation is that a compact object should have formed at the time of explosion.
Despite prompt neutrino emission observed at the burst \citep{Bionta1987,Hirata1987,Alekseev1987} indicating the presence of a neutron star (NS; \citealp{Burrows1988,Sukhbold2016}), the search for a compact object associated with SN~1987A has been the Grail of a long and continuous search:
All the observational searches have thus far proven unfruitful (e.g., \citealp{Manchester2007,Alp2018a,Zhang2018}). 

The detection of radio polarization by \cite{Zanardo2018} hints at the presence of magnetized shocks, potentially due to a compact object.
\cite{Alp2018a} proposed that a thermally-emitting NS could be dust-obscured, and that this may be detectable as a point source in far-IR or sub-mm images of the remnant.
Our MIRI images have not shown any evidence for such a point-like object.
%Although our MIRI images do not show any evidence of such a point-like object, what do sub-mm ALMA observations show?

\cite{Cigan2019} detected with ALMA a dust peak {\em `blob'}  that they attributed to either a pulsar wind nebula (PWN) or to a clump heated by ${}^{44}$Ti decay.
They argue that the most probable explanation is that the innermost region of dust and gas is heated by radiation from the NS, with an early development of a PWN, and propose that the identified central {\em `blob'} is due to warm ejecta heated by the NS.
Notwithstanding, they note that the ALMA data cannot disentangle whether the heating originates from grains heated directly by thermal X-rays from the NS, as in \cite{Alp2018a}, or by synchrotron radiation generated by the NS.

Figure \ref{fig:alma} shows the ALMA images at 315 and 679~GHz with contours from the MIRI images and the position of the warm {\em `blob'} found by \cite{Cigan2019}.
We, however, do not find any obvious correspondence between the position of this in the ALMA images and any enhancement in the MIRI images. 

In \cite{Fransson2023} a strong point source at the center of SN~1987A is found at 6.9861~$\mu$m, which is identified with the [\ion{Ar}{2}]  6.9853~$\mu$m line, blueshifted by $\sim 253$~km\, s$^{-1}$.
In addition, weaker lines from [\ion{Ar}{6}] 4.529~$\mu$m, [\ion{Ar}{3}] 8.991~$\mu$m, [\ion{S}{4}] 10.51~$\mu$m, and [\ion{S}{3}] 18.71~$\mu$m are identified.
The position, the low radial velocity, and the fact that this source is only seen in lines of highly-ionized S and Ar, show that the emission is originating in the explosive oxygen burning zone, dominated by Si, S, Ar and Ca, and that the ionizing source is likely to be the central NS in SN 1987A.
The exact nature of this is not yet clear.
Candidates are the thermal emission from the cooling NS, the non-thermal emission from a PWN, the shock from the expending PWN bubble, or a combination of these.
In any case, the presence of a compact object is strongly indicated.
With this background it is therefore natural to look for a corresponding point source in the MIRI images. 

Unfortunately, neither the F560W nor the F1000W filters cover the [\ion{Ar}{2}] 6.985~$\mu$m line, which is by far the strongest of the above lines and in addition is less contaminated by the strong dust continuum background at the longer wavelengths.
The other lines have fluxes $\la 4$~per~cent of this line and
%, with the exception of the [\ion{Ar}{6}] 4.529~$\mu$m line, 
are dominated by the dust background in the filters.
It is therefore not surprising that no point source corresponding to these lines are seen in the observed filters.
Imaging in the F770W filter, which contains the [\ion{Ar}{2}] 6.985~$\mu$m line, would be the most promising MIRI filter to reveal more details about the central emission source.

\begin{figure*} % Figure 15
\begin{center}
\includegraphics[width=\textwidth]{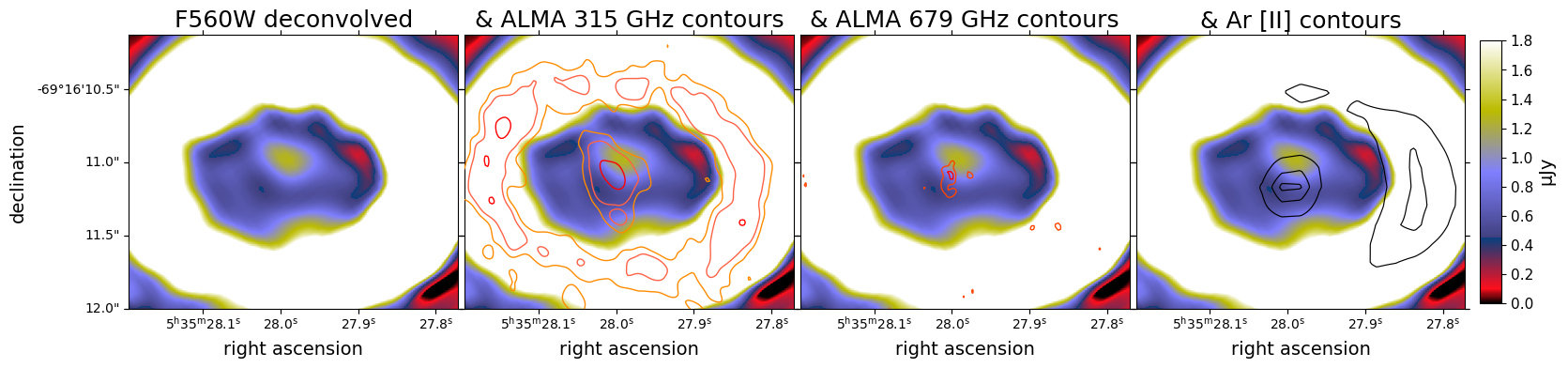} 
\end{center}
\caption{MIRI F560W image, deconvolved with the Richardson-Lucy algorithm. The color scale was set to highlight the faint ejecta.
The second, third and fourth panels from the left show this image together with contours from the ALMA 315 GHz image, the ALMA 679 GHz image, and the MRS [\ion{Ar}{2}] 6.9853~$\mu$m line, respectively. Note that the dust {\em `blob'} that \cite{Cigan2019} attribute to heating by the compact object can be seen in the third panel.}
\label{fig:f560w_ejecta}
\end{figure*}

The F560W image shows a detection of emission from the ejecta.
To highlight the structure within the ejecta more clearly we deconvoled the image with the Richardson-Lucy algorithm, using the WebbPSF model.
The result is shown in Figure~\ref{fig:f560w_ejecta}, showing an emission region just north of the center and a curved region of lower surface brightness just below it.
This morphology is very similar to that observed in the NIRCam F323N and F356W filters \citep{Arendt2023}, which are dominated by continuum emission in the ejecta region (see spectra in \citealt{Larsson2023}).
Figure~\ref{fig:f560w_ejecta} also includes contours from the ALMA 315 GHz image which probes cold dust, the dust {\em `blob'} at 679 GHz that \cite{Cigan2019} attribute to heating by the compact object, and the MRS [\ion{Ar}{2}] 6.9853~$\mu$m line associated with the compact object.
The [\ion{Ar}{2}] emission is centred south of the emission region in the F560W image. 

\subsection{Emission outside the ER} % Section 4.2
\label{sec:outside the ER}

In the MIRI images the ER is not as well defined in the mid-IR now as it was in past years \citep{Bouchet2006}.
Instead, extended emission outside but adjacent to the ER is seen, whose size (width of ellipse) increases with wavelength.
The images at shorter wavelengths have been blurred to the spatial resolution of the following longer wavelength images, in order to show, via a recursive subtraction between adjacent filters, a comparison of the flux evolution with wavelength.
The results are shown in Figure~\ref{fig:troisdiff}, which illustrates the increasing flux of the SW region with longer wavelengths, and also extension beyond the ER.
These images are a good indication of the E-W temperature variation in the ER itself.
Even though the inner ejecta are not directly seen in the individual images, this seems to indicate that they have a redder spectrum (possibly from [\ion{Fe}{2}] emission) than the ER.

To deconvolve, we instead used a modified version of the CLEAN method \citep{Hogbom1974} which looks for positive sources in the central region of the maps.

We also extracted templates from the images without the low level extensions, then convolved those templates with the PSF and obtained the same kind of extensions.

It can be seen that at the longest wavelength (25.5~\micron), the mid-IR emission apparently extends outside the known ER.
Note that an extension beyond the knotty ER has been seen in the HST data for some years now \citep{Fransson2015}.
The apparent extension in our MIRI images is however mostly the product of a PSF effect.
Unfortunately there is no PSF available of sufficient quality at these wavelengths to make a reliable correction.

\begin{figure*}[th] % Figure 16
  \includegraphics[width=9cm]{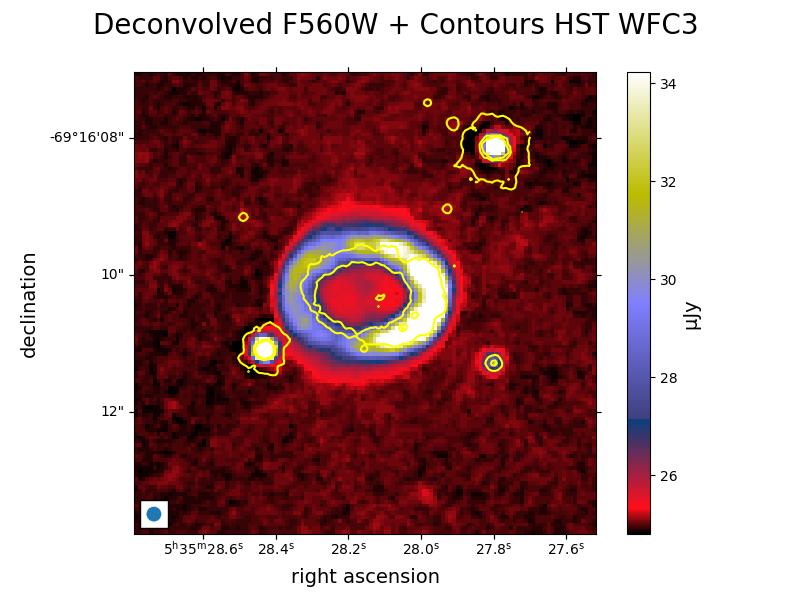}
  \includegraphics[width=9cm]{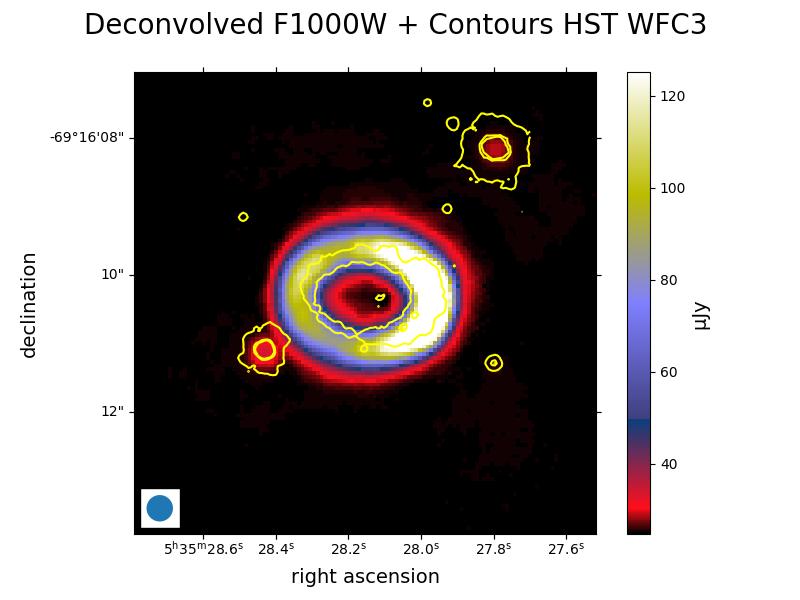}
  \includegraphics[width=9cm]{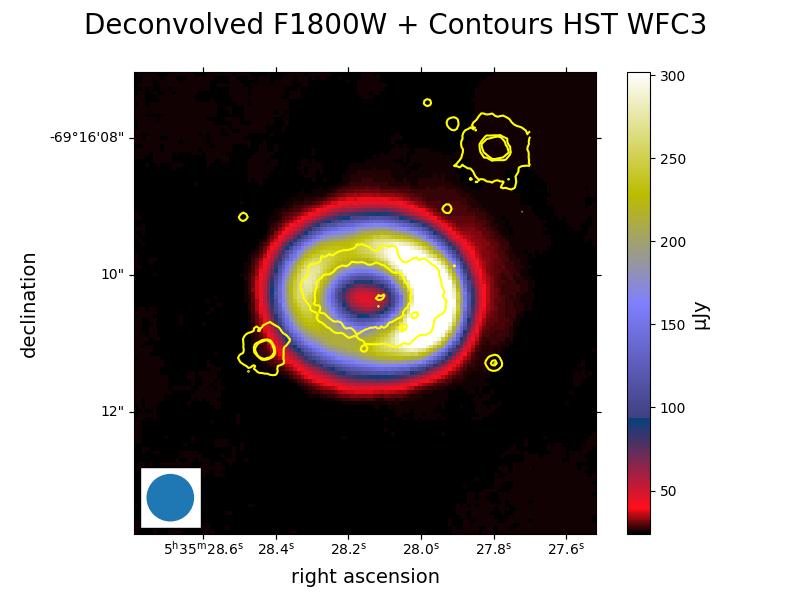}
  \includegraphics[width=9cm]{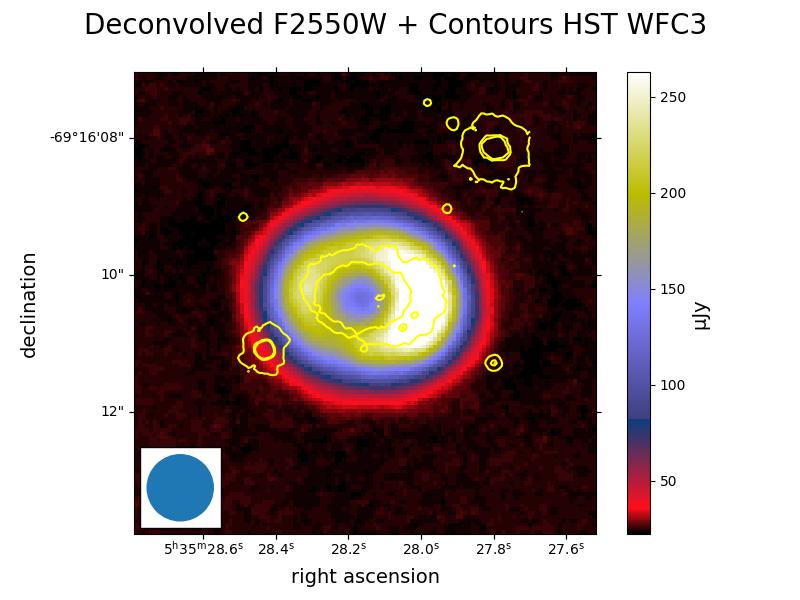}
  \caption{Deconvolved images by CLEAN method at 5.6, 10, 18, \& 25.5 $\mu$ using Web PSF as dirty beams and the corresponding clean beam (2D Gaussian fit) (FWHM Gaussian fit of the actual PSF for each wavelength are on lower left in each panel). Superimposed contours are based on the HST WCF3 image at $\lambda$  =0.502 $\mu$m. Most of the extensions around the ER are removed by deconvolution. Contours levels: [0.06, 0.6, 1 electrons/s]}
  %Deconvolved images by CLEAN %method at 5.6, 10, 18, \& 25.5 %$\mu$ per pixel, with the %corresponding clean beam (%Gaussian fit of the actual PSF %for each wavelength; lower %left in each panel).
  %Most of the extensions around %the ER are removed by %deconvolution.
  %Contours levels: [0.06, 0.6, 1 %electrons/s]}
  \label{fig:clean4images}
\end{figure*}

Results from our deconvolution of the MIRI images are presented in Figure \ref{fig:clean4images}, which shows clearly that the extension of the mid-IR emission seen in Figure \ref{fig:4images} may not be real.
In Figure \ref{fig:4images} we see a red ``skirt" growing with wavelength.
For the F2550W filter we find some structure in the red skirt.
But after our deconvolution using H\"{o}gbom's CLEAN algorithm, the structure vanishes (Fig.~\ref{fig:clean4images}).
Note that the CLEAN algorithm does not enhance the resolution, but replaces the dirty beam (with structure) by a ``cleaned" beam fitted by a 2D Gaussian.
We clearly see that the deconvolved F2550W image and the F560W image both contain a ring with a thickness of a few pixels.

When high-quality PSFs and efficient removal of odd/even row effects become available it will be possible to redo this study with higher confidence but for now we consider that no evidence of extensions have been shown in these broad-band images.
We note, however, that evidence for extensions has been seen in MRS spectral channels that isolate fine-structure line emission \citep{Jones2023}. 

Finally, no hint of the two outer rings can be seen in any of our four MIRI broad-band images, although forbidden-line emission from the outer rings is seen at several wavelengths in the much higher resolving power MRS spectral images of SN~1987A \citep{Jones2023} {\bf} where we could only see the outer rings in specific emission lines but not in dust continuum emission. We would therefore expect the NIRCam images reported by \cite{Arendt2023} showing the outer rings to be those dominated by line emission. Of the eight filter images shown in Fig.1 of \cite{Arendt2023}, two show the outer rings: the F164N ([Fe II]+[Si I]); and F405N (Br-alpha) images.

\section{Summary and conclusions} % Section 5
\label{sec:conclusions}

Our main results can be summarized as follows:

\begin{itemize}
 \item We have obtained MIRI images of SN~1987A taken with four different filters: F560W, F1000W, F1800W and F2550W, each covering a 56.3$\times$56.3~arcsec field of view. The images are dominated by the bright $\sim$2~arcsec diameter ER at their center but show extended nebulosity towards the edges of the field, around a cavity with an angular diameter of $\sim$30~arcsec that surrounds SN~1987A.

 \item Thermal dust emission dominates the observed filter  in-band fluxes. A comparison of the filter profiles with the MIRI-MRS 5-28~$\mu$m spectrum of the ER presented by \cite{Jones2023} shows that line emission makes only a small contribution to the total flux seen in each filter (ranging from $<$0.1 per cent up to 2.9~per~cent, depending on the filter). Excellent agreement was found between the integrated fluxes measured for the ER with each filter and the equivalent MRS photometric fluxes.

 \item Spatial dust temperature and mass maps were constructed for the region encompassing the ER. Using a silicate and amorphous carbon model fit to the data from all four filters, dust temperatures vary from 120~K to 165~K and masses from $0.2$ to $1.5~10^{-8}M_\sun$ per pixel, while a fit using the {\em astrodust} mixture of \cite{Hensley2023} led to slightly higher dust temperatures and masses from $0.5$ to $3.5~10^{-8}M_\sun$ per pixel. The total mass is $1.3~10^{-5}M_\sun$ and $2.8~10^{-5}M_\sun$, respectively. The total mass is 10 times greater than the mass reported in 2006 \citep{Bouchet2006}. A fit with the {\em astrodust} model to just the 18-$\mu$m to 25.5-$\mu$m image ratio led to lower dust temperatures of between 105~K and 140~K.

 \item No evidence for an unresolved compact object was found in any of our
 four continuum-dominated images.

 \item We searched for evidence for extended emission beyond the `knotty' ER seen at shorter wavelengths - however most of the apparent extensions around the ER seen in our MIRI images appear to arise from PSF effects and were removed by deconvolution with the CLEAN algorithm.
\end{itemize}

The most striking fact is that in our temperature maps the inner ring (ER) is not well defined.
The IR emission in the temperature maps extendswell beyond the ER, which is difficult to visualize in our direct images.
The other highlight is that the IR emission from the east side of the ring is quite a bit fainter at these mid-IR wavelengths than in the west side.
This seems to be an indication that dust has been disrupted in the east region.
On both sides, we may see remains of silicates.
This suggests that it is now difficult to use conventional models to model the IR emission in and outside the ER.
Although this is a hazardous and very speculative hypothesis, we suggest that the emission from beyond the ER may be synchrotron or another emission mechanism (on-going work).

\section*{Acknowledgements}
This work is based on observations made with the NASA/ESA/CSA James Webb Space Telescope.
{\bf} All of the data presented in this article were obtained from the Mikulski Archive for Space Telescopes (MAST) at the Space Telescope Science Institute which is operated by the Association of Universities for Research in Astronomy, Inc., under NASA contract NAS 5-03127 for JWST. The specific observations analyzed can be accessed via \dataset[DOI: 10.17909/k6j3-vm72]{https://doi.org/10.17909/k6j3-vm72}.
These observations are associated with program \#1232.
LL acknowledges support from the NSF through grant 2054178.
OCJ acknowledges support from an STFC Webb fellowship. 
CN acknowledges the support of an STFC studentship.
MM and NH acknowledge that a portion of their research was carried out at the Jet Propulsion Laboratory, California Institute of Technology, under a contract with the National Aeronautics and Space Administration (80NM0018D0004).
MM and NH acknowledge support through NASA/JWST grant 80NSSC22K0025.
PJK and JJ acknowledge support from the Science Foundation Ireland/Irish Research Council Pathway programme under Grant Number 21/PATH-S/9360.
MJB acknowledges support from European Research Council Advanced Grant 694520 SNDUST. 
ON acknowledges the NASA Postdoctoral Program at NASA Goddard Space Flight Center, administered by Oak Ridge Associated Universities under contract with NASA.
ASH is supported in part by an STScI Postdoctoral Fellowship.
LC acknowledges support by grant PIB2021-127718NB-100, from the Spanish Ministry of Science and Innovation/State Agency of Research MCIN/AEI/10.13039/50110001103.
CF and JL acknowledges support from the Swedish Space Agency. TT acknowledges financial support from the UK Science and Technology Facilities Council and the UK Space Agency. JB and BV thank the Belgian Federal Science Policy Office (BELSPO) for the provision of financial support in the framework of the PRODEX Programme of the European Space Agency (ESA). Finally, we thank the referee for his comments and constructive advices.

\facilities{JWST (MIRI)}

\software{Astropy \citep{Astropy2022},
  Matplotlib \citep{Hunter2007},
  PyPHER \cite{Boucaud2016},
  GDL \citep{Park2022}, MPFIT \citep{Markwardt2009}}

\bibliography{Biblio/sn87a_refs,Biblio/add2bib.bib} %biblio_sn1987A}{}
\bibliographystyle{aastex631/aasjournal}

\appendix

\section{Reduction Method} % Section A
\label{sec:Appendix}

We did not follow the architecture of the official pipeline (see \url{https://jwst-docs.stsci.edu/jwst-science-calibration-pipeline-overview)} for the following reasons:

\begin{itemize}

\item Not all the configuration files were up-to-date, and the pipeline was not fully tested at the time of our analysis.
  
\item We were mainly interested in a small region around the center (40 pixels wide), encompassing the ER and the outer rings.
Note that in our dedicated pipeline, we only used three calibration files for (i) the non-linearity correction, (ii) the mask for the dead and hot pixels, (iii) the photometric conversion to Janskys.
The dark correction was included in the correction of the background.
The flat field correction was small, because the region of interest is small and in the center of the image.
The mosaicking took care of that correction.

\end{itemize}

\begin{figure} % Figure 17
  \includegraphics[width=8.cm]{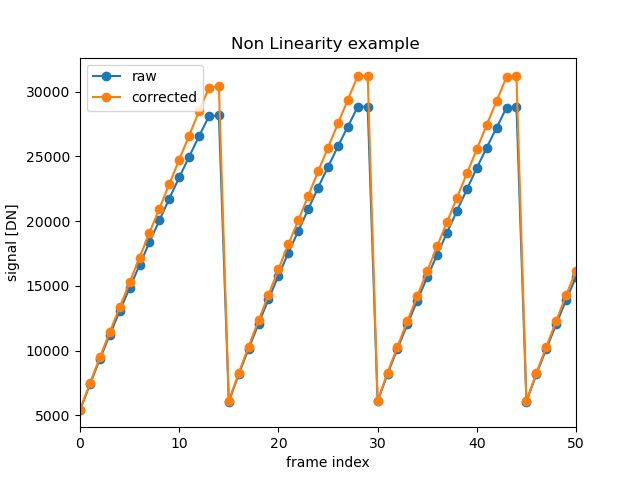}
  \caption{Raw data sample ramps, before and after non-linearity correction.}
  \label{fig:rampes4a}
\end{figure}

\begin{figure} % Figure 18
  \includegraphics[width=8.cm]{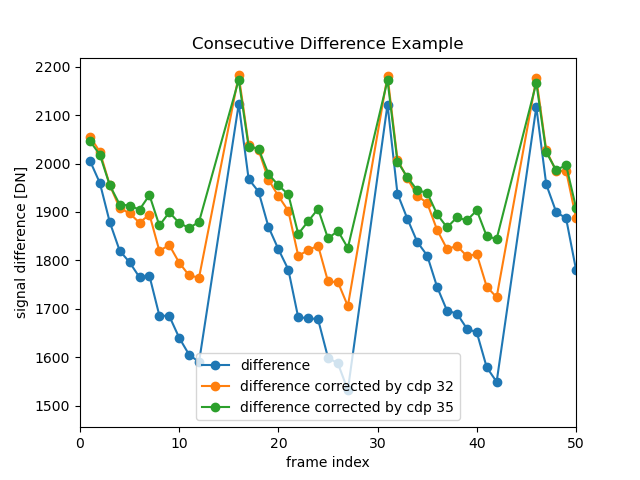}
  \caption{Differences of two consecutive frames, with and without linearity correction.
  With a perfect correction the differences should be constant, which is clearly not the case.
  However, there is a periodic pattern which clearly shows that it can be improved.}
  \label{fig:rampes4b}
\end{figure}

\begin{figure}[!th] % Figure 19
  \includegraphics[width=8.5cm]{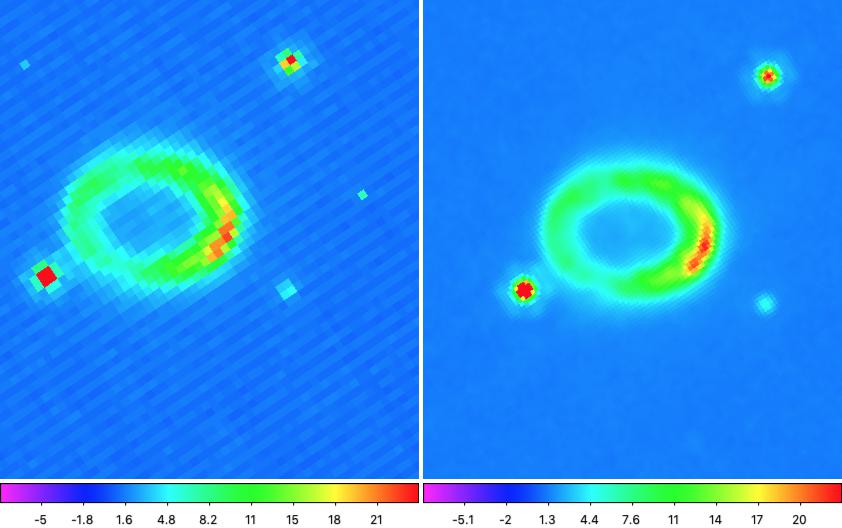}
  \caption{Background subtraction for the 5.6 \micron\ filter, odd/even effect correction: without correction (left panel), with correction (right panel).}
  \label{fig:vignettes}
\end{figure}

We thus proceeded as follows: 
\begin{enumerate}
\item We applied a non-linearity correction (see Figure~\ref{fig:rampes4a}).
The ramps were then transformed into slopes (linear regression with deglitching of the official pipeline), and we computed the median of consecutive differences (see Figure~\ref{fig:rampes4b}).
  
\item We computed the differences of two sequential consecutive frames.
Outliers were thus flagged (deglitching), and the first and the two penultimate frames were disregarded.
  
\item We computed the background outside of the ROI (Region of Interest) and interpolated on the overall image, to be subtracted.
This allowed the (weak) dark current to be taken into account, and to correct for the odd-even row effect (Figure~\ref{fig:vignettes}).

\item We used the median of the sequential consecutive differences to compute the slopes.

\item We merged the different dithered pointings for each filter.
  
\item We then calibrated the data in Jansky pixel$^{-1}$ units.
 
\item We use the WCS alignment tool JHAT \citep{Rest2023_jhat} to align the F560W and F1000W images to Gaia DR2, with a scatter of about 0.1 pixels. In addition, the F1000W shows a small systematic bias of about 0.2 pixels across the full detector.
The longer wavelength images do not have enough stars to do a reliable WCS alignment, and we therefore propagate the WCS solution from F560W and F1000W to F1800W and F2550W.
The next step was to check the astrometry with companion stars 2 and 3, seen at 5.6~\micron{} and marginally at 10~\micron.
In the dithering mode, the official pipeline gives relative xoffset and yoffset and absolute WCS values.
These values are half-integers, so we had to re-bin by a factor of 2 (centering and mean).
Note for example, that comparing filters F560W and F1000W we see a one pixel shift, which is most probably due to the fact that different filters introduce slight deflections of the beam.

\end{enumerate}

\end{document}